\RequirePackage{fix-cm}
\documentclass[smallextended]{svjour3}       

\usepackage{latexsym}
\usepackage{bm}
\usepackage{epsfig,amsopn}
\usepackage{graphicx}
\usepackage{amsmath,amssymb}
\usepackage{braket}
\usepackage{comment}
\usepackage{bm}
\usepackage{mathptmx}
\usepackage{enumerate}
\usepackage{epsfig,amsopn}
\usepackage{ulem}
\usepackage{graphicx}
\usepackage{amsmath,amssymb}
\usepackage{braket}
\usepackage{comment}
\usepackage{xcolor}

\newcommand{\nn}{\nonumber}

\newcommand{\eref}[1]{Eq.~(\ref{#1})}%
\newcommand{\fref}[1]{Fig.~\ref{#1}} %
\newcommand{\Fref}[1]{Figure~\ref{#1}}%
\newcommand{\sref}[1]{Sec.~\ref{#1}}%
\newcommand{\be}{\begin{equation}}
\newcommand{\ee}{\end{equation}}
\newcommand{\bea}{\begin{eqnarray}}
\newcommand{\eea}{\end{eqnarray}}

 \journalname{Journal of Statistical Physics}

\begin{document}

\title{Asymptotic behavior of the velocity distribution of driven inelastic gas with scalar velocities:  analytical results}

\titlerunning{Velocity distribution in driven one-dimensional granular gases}  

\author{V. V. Prasad  \and
        R. Rajesh 
}

\institute{V. V. Prasad \at
              Department of Physics of Complex Systems, Weizmann Institute of Science, Rehovot 7610001, Israel\\
              The Institute of Mathematical Sciences, C.I.T. Campus, Taramani, Chennai-600113, India\\
              Homi Bhabha National Institute, Training School Complex, Anushakti Nagar, Mumbai-400094, India\\
              \email{prasad.vv@weizmann.ac.il}         
           \and
           R. Rajesh \at
             The Institute of Mathematical Sciences, C.I.T. Campus, Taramani, Chennai-600113, India\\
             Homi Bhabha National Institute, Training School Complex, Anushakti Nagar, Mumbai-400094, India\\
             \email{rrajesh@imsc.res.in} 
             }

\date{Received: date  / Accepted: date}

\maketitle

\begin{abstract}

We determine the asymptotic behavior of the tails of the steady state velocity
distribution of a homogeneously driven granular gas comprising of particles having a scalar velocity.
A pair of particles undergo binary inelastic collisions at a rate that is proportional to a power of their relative
velocity.  At constant rate, each particle is driven by multiplying its velocity by a factor $-r_w$ and  adding a stochastic noise.  
When $r_w <1$, we show 
analytically that the tails of the velocity  distribution are primarily determined by the noise statistics, and 
determine analytically all the parameters characterizing the velocity distribution
in terms of the  parameters characterizing the stochastic noise.  Surprisingly, we find
logarithmic corrections to the leading stretched exponential behavior.  When $r_w=1$, we show that
for a range of distributions of the noise, inter-particle collisions lead to a universal tail for the velocity distribution.

\end{abstract}

\section{Introduction}

Granular systems are ubiquitous in nature. Examples include sand, powders, grains, interstellar objects like asteroids and so on.
Composed of macroscopic particles, the combination of inelastic collisions and external driving
results in granular systems exhibiting complex phenomena such as size segregation, 
pattern formation, jamming, memory 
effects, shocks, etc.~\cite{Jaeger:96,Aranson:06,Goldhirsch:93,Li:03,Corwin:05,Prados:14,Lasanta:17,joy2017shock}.
Being far from equilibrium, a general theory to understand granular systems has been lacking.
However, a class of systems for which theoretical progress has been possible are  dilute granular 
matter, also  known as granular gases. In addition to exhibiting complex macroscopic behavior, granular
gases have been a testing ground for both kinetic theory as well as for simple solvable models.  

When isolated, a portion of the kinetic energy of the granular gas is dissipated in each collision, and the constituent particles
eventually form clusters leading to strong spatial inhomogeneities~\cite{Goldhirsch:93,Brey:96,Esipov:97,Ben-naim:99,Ben-naim:00,Nie:02,Supravat:11,Pathak:14a,Pathak:14,shinde2007violation,Subhajit:14,Subhajit:17,brilliantov2018increasing}. 
However, when driven homogeneously the system reaches a steady state that is homogeneous in space~\cite{Vanzon:04,Scholz:17}.  
The single particle velocity distribution $P(v)$~[`$v$' is the magnitude  of the velocity which is in general a vector] of such driven granular gases has been the subject of many
different studies. Of particular interest is the question of whether the asymptotic behaviour of $P(v)$ is universal, and if yes then what its form is. This question has been addressed in several experiments, numerical simulations, and within kinetic theory. A recent review may be found in Ref.~\cite{windows2017granular}. 
Experimental systems of driven granular gases typically
consists of collections of granular particles, such
as steel, glass beads,  etc., that may be spherical, ellipsoidal, dumbbell-shaped, etc.,  that undergo inelastic collisions and is driven 
either through collision of the particles with vibrating walls~\cite{Clement:91,Warr:95,Kudrolli:97,Olafsen:98,Olafsen:99,Losert:99,Kudrolli:00,Rouyer:00,Blair:01,Vanzonexpt:04,reis2007forcing,wang2009particle,Scholz:17,vilquin2018shock,wildman2009granular} or as bilayers where the vibrated bottom layer drives the top layer~\cite{Baxter:03,baxter2007temperature,windows2013boltzmann}, or by using electric~\cite{Aranson:02,Kohlstedt:05} or magnetic fields~\cite{Schmick:08,Falcon:2013}.  In some experiments, the effects of gravity are minimised by performing them in  microgravity~\cite{tatsumi2009experimental,hou2008velocity,grasselli2015translational}.
Several experiments observe a universal stretched exponential form $P(v)\sim\exp(-a|v|^\beta)$ with 
$\beta\approx 1.5$  for wide range of system parameters~\cite{Losert:99,Rouyer:00,Aranson:02,reis2007forcing,wang2009particle,tatsumi2009experimental,Scholz:17,vilquin2018shock}, while 
other experiments find that $\beta$ differs from $1.5$ and lies between  $1$ and $2$ or is a gaussian, and may
depend on the driving parameters~\cite{Olafsen:99,Kudrolli:00,Blair:01,baxter2007temperature,Schmick:08,hou2008velocity,wildman2009granular,Falcon:2013,windows2013boltzmann,grasselli2015translational}.
Numerical simulations~\cite{puglisi1998clustering,puglisi1999kinetic,Moon:01,Vanzon:04,Vanzon:05,Cafiero:2002,burdeau2009quasi,gayen2008orientational,gayen2011effect,rui2011velocity,Prosendas:18,kang2010granular} have also been inconclusive. The velocity distribution for a one dimensional gas driven through a thermal bath is observed to have a gaussian form in the quasi-elastic limit but deviates from gaussian when the collisions are inelastic~\cite{puglisi1998clustering,puglisi1999kinetic}.
For a three dimensional granular gas driven homogeneously with a momentum conserving noise,  it was shown that 
$\beta \approx 1.5$ for large inelasticity, whereas 
$\beta$ approaches $2$ when collisions are near-elastic~\cite{Moon:01}. Also, when the granular gas is polydispersed, one finds a range of $\beta$~\cite{rui2011velocity}. 
Similar study on a bounded two dimensional granular system 
observes $\beta \approx 2$ for a  range of coefficient of restitution and density~\cite{Vanzon:04,Vanzon:05}, while a two dimensional
system driven through the rotational degrees of freedom find $\beta \approx 1.42$~\cite{Cafiero:2002}. Molecular dynamics simulations of a uniformly heated granular gas with solid friction find
$\beta=2$~\cite{Prosendas:18}. Sheared granular gases have also been studied via simulations, which find $\beta \approx 1.5$~\cite{gayen2008orientational,gayen2011effect}, while those for bilayers show $\beta =2$~\cite{burdeau2009quasi}.
Models with extremal driving have also been studied which exhibit intermediate power law behaviour~\cite{kang2010granular}. 
Finding the asymptotic behavior of the velocity statistics from experiments and direct simulations is 
challenging due to poor sampling near the tails and the presence of strong crossovers from the behaviour of the distribution at 
small velocities to the asymptotic behaviour at high velocities, making it difficult to unambiguously determine $\beta$.

In the homogeneous limit, the analysis may be simplified by integrating out or ignoring the spatial degrees of freedom. 
The velocity distribution may then be studied either 
using kinetic theory~\cite{Brilliantov:04} where 
molecular chaos is assumed to break  BBGKY hierarchy, or by analyzing simple microscopic models like the inelastic 
Maxwell model~\cite{Bobylev:00,Ben-naim:00},
where the collision rate between particles is assumed to be the same for all pairs of particles. Within kinetic theory, the tail
of the velocity distribution may be determined by linearizing the Boltzmann equation~\cite{Noije:98}. When the rate
of collision of a pair of particles is proportional to their relative 
velocities and  the driving is described by a diffusive term, it was shown that the velocity distribution for large velocities $|v| \gg v_{rms}$, is a stretched exponential with exponent $\beta=3/2$~\cite{Noije:98}, independent of spatial dimension. 
In Maxwell gases, most of the rigorous results are for systems with scalar (one dimensional) velocities, as the collisions 
in 2 and 3 dimensions are more complicated.
Maxwell gases with scalar velocities, with constant rate of collision and diffusive driving ($d{v}/dt={\eta}$ where ${\eta}$ is gaussian white noise), may be solved exactly to yield an exponential decay for the velocity distribution
with $\beta=1$~\cite{Ernst:02,Antal:02,Santos:03}. 
Both these cases can be analysed using single framework by considering
a general model where a pair of particles with
relative velocity $\vec{v}$ collide with rate $|\vec{v}|^\delta$~\cite{Ernst2003,ernst2006rich,ernst2006,barrat2007quasi}.
Even though one is especially interested in physical limit of $\delta=1$~(also called {\it{hard-spheres}}),  
the general model is useful for more general kernels which  may phenomenologically have other values of $\delta$~\cite{Kohlstedt:05}.
Note also, that in elastic kinetic theory $\delta=-3$ corresponds to Coulomb interaction.
For the model with collision rate $|\vec{v}|^\delta$ the Boltzmann equation with diffusive driving may be 
analyzed to yield $\beta= (2+\delta)/2$~\cite{Ernst2003} retrieving the special limits of hardsphere~($\delta=1$) and Maxwell~($\delta=1$) gases. 
For a review on results obtained from kinetic theory, see Ref.~\cite{Villani:06}.

Within the molecular chaos hypothesis, the main issue is how to model noise. While the tails may be determined for random acceleration (diffusive driving)
the system   does not reach a steady state, as the velocity of the center of mass does a random walk leading
to the linear divergence of total energy with time~\cite{Prasad:13}. Hence the steady state solutions for diffusive driving is valid only on the center of mass frame.   A more realistic driving scheme is  dissipative driving, in which when a particle 
with velocity $\vec{v}$ is driven, its velocity is modified to  $\vec{v}^\prime=-r_w \vec{v}+{{\bf\eta}}$, where $\vec{v}'$ is the new velocity, $r_w$ is  a parameter by which the velocity is dissipated and 
${\bf{\eta}}$ is uncorrelated random noise taken from a distribution $\Phi({\bf{\eta}})$~\cite{Prasad:13}. For $r_w \in (-1,1]$, the system reaches
a steady state at long times~\cite{Prasad:13}. The  model with scalar velocities and  dissipative driving has been recently studied in detail~\cite{Prasad:17} 
for noise distributions that has the asymptotic behavior $\Phi(\eta)\sim \exp(-b|\eta|^\gamma)$ for $\eta^2 \gg \langle\eta^2\rangle_\Phi$, where the average  $\langle..\rangle_\Phi$ refers to averaging over the noise distribution $\Phi(\eta)$. By studying
the moment equations numerically~\cite{Prasad:17}, it was shown that  when $r_w < 1$, 
the velocity distribution~$P(v)\sim \exp(-a|v|^\beta)$, is non-universal 
with $\beta = \gamma$, while when $r_w=1$, 
$\beta = \min(\gamma,1)$. Thus, $\beta$ is universal (independent of noise) only when  $r_w=1$  and $\gamma \ge 1$.  One could also make predictions about the coefficient $a$.
It was observed numerically that $a=b$ for $\gamma < 1$ and $a\neq b$ for $\gamma >1$.  All these results were based
on numerical analysis and restricted to the case $\delta=0$, corresponding to the Maxwell gas. The
exact form of $a$ is known for the Maxwell gas only for two cases: one, for $r_w < 1$ and  $\gamma=2$  when $a=b(1-r_w^2)$~\cite{Prasad:13}, and other, for $r_w=1$ and $\gamma\geq1$, when $a=\omega^*$, where $\omega^*$ is the solution for $1=\lambda_d f(\omega)$, with $\lambda_d$, the rate of driving and $f(\omega)=\langle e^{-i\omega\eta}\rangle_{\Phi}$, the characteristic function of the noise distribution~\cite{Prasad:17}.  We note that, for both diffusive driving as well as dissipative driving, all exact results have been limited to the case $\delta=0$.

In this paper,  we choose to study a system with scalar velocities as the analysis there is much simpler (Study of 
an inelastic gas with two-dimensional velocities may be found in Ref.~\cite{Prasad:19}). 
For the  inelastic gas with scalar velocities, driven homogeneously through dissipative driving, 
we carry out an exact analysis of the equations satisfied by the moments of the velocity for a general collision kernel, corresponding to arbitrary $\delta$.   To our knowledge, this is the only solution that exists for non-zero $\delta$. We determine the
different parameters of the velocity distribution  in terms of the asymptotic behavior of the  noise distribution $\Phi(\eta)$ that is characterized by the 
parameters $\gamma$, $b$, and  ${\tilde{\chi}}$ through $\Phi(\eta) \sim |\eta|^{\tilde{\chi}} \exp(-b|\eta|^\gamma)$ for large $|\eta|$.
Surprisingly, we find the presence
of logarithmic corrections to the leading stretched exponential decay even in the absence of any such corrections in $\Phi(\eta)$. In addition, the distribution depends strongly on  whether $|r_w|<1$ or $r_w=1$. In particular, 
for large velocities ($v^2 \gg \langle v^2 \rangle$), we assume that the leading behaviour for the velocity distribution is a stretched exponential. We show that, within this assumption, a self-consistent solution for the equations satisfied by the moments of the velocity may be found. Though we are unable to show the uniqueness of our self-consistent solution, we believe that the  results, thus obtained, are exact. Our results are summarised below. We show that asymptotically, the velocity distribution decays as
\be
\label{velocity pdf with log correction 1}
\ln P(v) = -a|v|^\beta-a' (\ln|v|)^2 + \chi \ln |v|+ \ldots,~~|r_w|<1, 
\ee
where
\begin{align}
\begin{split}
\label{velocity pdf with log correction 2}
\beta=& \gamma, \\
a=& \begin{cases}
b,  \qquad \qquad \qquad \quad \gamma \leq 1,\\
b\left[1-r_w^\frac{\gamma}{\gamma-1}\right]^{\gamma-1},  ~\gamma > 1,\\
\end{cases}\\
a'=&\begin{cases}
0,  \qquad \qquad \qquad \qquad ~~ \gamma \leq 1,\\
\frac{\gamma(\gamma-1)}{2\ln(r_w)}\left[\frac{1}{\gamma}+\frac{\tilde{\chi}}{\gamma}-\frac{\tilde{\delta}}{\gamma}-\frac{1}{2}\right], ~\gamma > 1,\\
\end{cases}\\
\chi=& \begin{cases}\tilde{\chi}-\tilde{\delta}, ~~\gamma \leq1,\\
0, ~~~~~~~~~\gamma >1
\end{cases}
\end{split}
\end{align}
where $\tilde{\delta}=\max[\delta,0]$.
For $r_w=1$,  we show that when $\delta \leq 0$,
\begin{align}
\label{velocity pdf with log correction 3}
\ln P(v) = -a|v|^\beta + \ldots,~~r_w=1,~-2<\delta\leq 0,
\end{align}
where
\begin{align}
\beta =&\min\left[\frac{2 + \delta}{2},\gamma \right], \quad r_w=1,~ -2<\delta\leq 0,\\
a=& \begin{cases}
 \frac{4}{2+\delta}\sqrt{\frac{\lambda_d}{N_2 \lambda_c}},  ~-2<\delta < 0,~~\frac{2+\delta}{2}<\gamma,\\
 \omega^* ,   \qquad \qquad ~~~~~ \delta = 0,\qquad~~~~1\leq\gamma,\\
b, \qquad \qquad \quad  ~~~~\delta\leq 0,\qquad~\frac{2+\delta}{2}>\gamma,
\end{cases} 
\end{align}
where $N_2$ is the second moment of the noise distribution, and $\lambda_c$ and $\lambda_d$ are rates
for collision and driving respectively [see \sref{sec:model} for a precise definition of rates], and $\omega^*$ is the solution for 
$1=\lambda_d f(\omega)$, with $f(\omega)=\langle e^{-i\omega\eta}\rangle_{\Phi}$ being the characteristic function of the noise distribution~\cite{Prasad:17}.
For $r_w=1$ and $\delta > 0$, we show that
\begin{align}
\label{velocity pdf with log correction 4}
\ln P(v) = -a|v| (\ln|v|)^{\frac{\gamma-1}{\gamma}}+ \ldots, r_w=1,\delta > 0,\gamma >1,
\end{align}
where
\be
a=\left[\left(\frac{\delta\gamma}{\gamma-1}\right)^{\gamma-1}b\gamma \right]^\frac{1}{\gamma},r_w=1,\delta > 0,\gamma >1.
\label{eq:last}
\ee
 When $\gamma \leq 1$ and $\delta>0$, the velocity distribution is asymptotically same as that of the noise distribution.

The remainder of the paper is organized as follows. In Sec.~\ref{sec:model}, we define the model precisely. Section~\ref{sec:momentanalysis}
contains a detailed analysis of the equation satisfied by the various moments of the velocity distribution. This analysis allows us
to obtain a precise characterization of the asymptotic behavior of the distribution. The analytical results for $\delta=0$, thus obtained,
are compared with results from numerical analysis  in Sec.~\ref{sec:numerical}. Section.~\ref{sec:summary} contains a summary and 
discussion of the results.

\section{Model \label{sec:model}}

Consider $N$ particles, labeled $i=1, \ldots, N$, of equal mass, each having a scalar velocity denoted by $v_i$. 
The velocities evolve  through  both momentum-conserving inelastic collisions among particles as well as external driving. 
Particles $i$ and $j$ collide at a
rate $2 \lambda_c |v_i-v_j|^\delta /(N-1)$. The factor $2/(N-1)$ is included so that the total rate of collision of the $N(N-1)/2$ pairs is proportional
to $N$. For Maxwell gases the collision rate is independent of the relative velocity, i.e., $\delta=0$, 
while for hard sphere gases the collision rate is proportional to the relative velocity, i.e., $\delta=1$.
In a collision, momentum is conserved and  the velocities are modified according to
\begin{equation}
\begin{split}
\label{interparticle collision}
v_{i}' &=    (1-\alpha) v_i+  \alpha v_{j},\\ 
v_{j}'&=\alpha v_i + (1-\alpha) v_{j},
\end{split}
\end{equation} 
where the unprimed and primed ($\prime$) velocities refers to the pre- 
and post-collision velocities. Also, 
\be
\alpha=\frac{1+r}{2},
\ee
where $r\in[0,1)$ is the coefficient of restitution, such that $\alpha \in[1/2,1)$. 
The two limits $r=1$ ($\alpha=1$) and $r=0$ ($\alpha=1/2$) correspond to elastic and maximally dissipative collisions
respectively.

Each particle is driven at rate $\lambda_d$. When driven, the
new velocity of a particle is given by
\begin{equation}
\label{wall collision}
v_{i}^\prime =  -r_w  v_i + \eta,
\end{equation} 
where $r_w \in (-1,1]$  quantifies how the speed is reduced upon driving, and the noise  $\eta$ mimics 
stochasticity in driving. Though this form of driving was introduced earlier~\cite{Prasad:13}, we give below several motivations to chose this particular model of driving.  

Firstly, when $r_w\not=-1$ [\eref{wall collision}]  the system reaches a steady 
state at large times, and overcomes the drawback of diffusive driving ($r_w=-1$) for which there is no steady state.
For the Maxwell gas, this may be shown analytically by obtaining the exact time evolution of the total energy~\cite{Prasad:13}. For a general
collision kernel, the existence of steady state may be seen through Monte Carlo simulations for both the gas with one-dimensional velocities~\cite{Prasad:13}, as well as two-dimensional velocities~\cite{Prasad:19}. Secondly, the driving in the limit $r_w=1$ captures the scenario described by kinetic theory models with a diffusive term for driving, and thus
provides a rigorous check for its predictions of $\beta=3/2$. To understand this, let us examine  the evolution of the distribution function $P(v,t)$:
 \begin{align}
\frac{d P(v,t)}{dt}=&- 2 \lambda_c \int  dv_2 |v-v_2|^\delta P(v,t) P(v_2,t)\nn \\
&+2 \lambda_c \int \int dv_1  dv_2 |v_1-v_2|^\delta P(v_1,t) P(v_2,t) \delta\left[(1-\alpha) v_1+  \alpha v_{2} -v\right] \nonumber \\
&-\lambda_d P(v,t) +
\lambda_d \int \int d\eta  dv_1 \Phi(\eta) P(v_1,t) \delta\left[ -r_w v_1+\eta -v \right],
\label{eq:evolution}
\end{align}
where the the product measure for the joint distribution $P(v_1,v_2)= P(v_1) P(v_2)$ is invoked, due to lack of correlations between the velocities of 
different particles in the large system size limit, arising from the fact that pairs of particles collide at random.
Note that this is true only in the steady state and that for diffusive driving, the system does not reach a steady state due to diverging 
 correlations between particles~\cite{Prasad:13}.  The first two terms on the right hand side of \eref{eq:evolution}
describe the loss and gain terms due to inter-particle collisions. The third and fourth terms on the right hand side describe the loss and
gain terms due to driving. The driving terms may be analysed for small $\eta$ as follows. Let
\be
I_D= -\lambda_d P(v,t)
+\lambda_d \int \int d{\eta} dv_1 \Phi(\eta) P(v_1,t) \frac{1}{r_w}\delta\left[\frac{\eta -v}{r_w}- v_1 \right].
\ee
Integrating over $v_1$, and using the symmetry of the distribution $P(v) = P(-v)$, we obtain
\be
I_D= -\lambda_d P(v,t)
+\frac{\lambda_d}{r_w} \int \int d\eta dv \Phi(\eta) P\left(\frac{v -\eta}{r_w},t \right).
\label{eq:aaaaa}
\ee
Now setting $r_w=1$, and Taylor expanding the integrand about $\eta=0$, and then integrating over $\eta$, Eq.~(\ref{eq:aaaaa}) reduces to
\be
I_D= \frac{\lambda_d \langle \eta^2 \rangle}{2} \frac{\partial^2 P(v)}{\partial v^2} +\mathrm{ higher ~order ~terms}, \quad r_w=1.
\label{eq:aaaaaa}
\ee
When the higher order terms are ignored, the resulting equation for $P(v)$ for $r_w=1$ is the same as that was 
analysed in Ref.~\cite{Noije:98} to obtain the well-known result of $\ln P(v) \sim -v^{3/2}$. It is not apriori clear 
whether this truncation is valid. This is because  the tails of the velocity distribution could be affected by tails of the noise distribution, 
in which case higher order moments of noise may contribute.

Thirdly, the model of dissipative driving may also be motivated by experimental systems, where driving is usually 
through collisions with a massive wall. In such 
a  wall-particle collision event, the velocity of a particle $i$,  $v_i$ and that of the wall, $v_w$ change 
to $v_i^\prime$ and $v_w^\prime$ respectively according to  $(v_i^\prime-v_w^\prime)=-r_w(v_i-v_w)$, where $r_w$ is the 
coefficient of restitution for the wall-particle collision. Considering 
$v_w^\prime\approx v_w$, for a massive wall,  rearranging and replacing   $(1+r_w)v_w$ by a noise $\eta$ results in
\eref{wall collision}. Within this motivation, $r_w$  takes only positive values.
 Since the stochasticity does not arise from a sum of many events, there is no compelling reason 
 for the  distribution for noise $\Phi(\eta)$ to be a gaussian.  For example, in experiments with sinusoidally oscillating
wall driving the particles, if the collision times are assumed to be random, 
$\Phi(\eta) \sim (c^2-\eta^2)^{-1/2}$, with $\eta\in (-c,c)$~\cite{Prasad:19}. This corresponds to the asymptotic behaviour of the noise being
$\Phi(\eta) \sim \exp(-a |\eta|^\gamma)$ with   $\gamma=\infty$.
To incorporate the general case  where the  tails of the noise distribution may differ from a  gaussian, we could consider noise 
distributions which decay either as  a stretched exponential  or as a power law or faster than a stretched exponential 
(for example $\exp[-e^{|\eta|}]$). However, we find that the power law tail and distribution decaying faster than stretched 
exponential may be considered as special cases of the stretched exponential form~[$\gamma =0$ and $\gamma=\infty$ 
respectively in Eq.~(\ref{eq:noise-pdf})] for large values of $\eta$. For these reasons,
we consider a set of noise distributions $\Phi(\eta)$ whose leading order asymptotic behaviour are stretched exponentials:
\be
\ln \Phi(\eta) \sim  -b|\eta|^\gamma  +\ldots+ \tilde{\chi} \ln|\eta|,~ b,\gamma>0, ~\mathrm{for~} \eta^2 \gg \langle \eta^2 \rangle_\Phi,
\label{eq:noise-pdf}
\ee 
where $\langle \cdots \rangle_\Phi$ denotes average with respect to $\Phi(\eta)$.

\section{\label{sec:momentanalysis} Analysis of moments for arbitrary $\delta$}

The equation obeyed by the different moments may be obtained by multiplying \eref{eq:evolution} by $v^{2 n}$ and integrating
over all velocities. In the steady state, after setting time derivatives to zero, we obtain
\begin{align}
\label{moment ratio delta non zero}
&2 \lambda_c [1-\alpha^{2n}-(1-\alpha)^{2n}]
\langle|v-v_1|^\delta v^{2n}\rangle
+\lambda_d (1-r_w^{2n})\langle v^{2n}\rangle
=\notag
\\
&2 \lambda_c \sum_{k=1}^{2 n-1} \binom{2n}{k}\alpha^{k}(1-\alpha)^{2n-k} \langle|v-v_1|^\delta  v^{2n-k}v_1^{k}\rangle\nn\\
&+\lambda_d\sum_{k=1}^{n}\binom{2n}{2k} r_w^{2n-2k} \langle v^{2n-2k}\rangle  N_{2 k}, ~n=1,2,3,\ldots,
\end{align} 
where 
$N_{2 k}=\langle \eta ^{2k}\rangle_\Phi $ is the $(2k)^{th}$ moment of the noise distribution, and $\langle \cdots \rangle$ denotes averaging
with respect to the steady  state velocity distribution. To clarify, in the case of joint moments such as $\langle|v-v_1|^\delta  v^{2n-k}v_1^{k}\rangle$ and $\langle|v-v_1|^\delta v^{2n}\rangle$, the averaging needs to be performed over the joint probability distribution 
$P(v,v_1)$.

We analyze \eref{moment ratio delta non zero} for large $n$. For large $n$, the left hand side of \eref{moment ratio delta non zero} is dominated
by the first term when $\delta>0$, and by the second term when $\delta<0$. 
Also, since $1/2\leq \alpha <1$, the terms $\alpha^{2 n}$ and $(1-\alpha)^{2 n}$ are negligibly small, and we may rewrite \eref{moment ratio delta non zero} as
\begin{align}
\label{moment-main}
&
\langle|v-v_1|^{\tilde{\delta}} v^{2n}\rangle
\sim
\sum_{k=1}^{n}\binom{2n}{2k} r_w^{2n-2k} \langle v^{2n-2k}\rangle  N_{2 k}+
\notag \\
&
\sum_{k=1}^{2 n-1} \binom{2n}{k}\alpha^{k}(1-\alpha)^{2n-k} \langle|v-v_1|^\delta  v^{2n-k}v_1^{k}\rangle, ~n \gg 1,
\end{align} 
where 
\begin{equation}
\tilde{\delta}=\begin{cases}
\max(\delta,0),&r_w<1, \\
\delta,& r_w=1,
\end{cases}
\end{equation}
 and $x\sim y$ means that $x/y = \mathcal{O}(1)$.
The first and second terms on the right hand side of \eref{moment-main} arise due to driving and  inter-particle collisions 
respectively.

We will assume that the velocity distribution is asymptotically a stretched exponential: 
\be
\ln P(v) = -a |v|^\beta + \Psi(|v|), ~v^2 \gg \langle v^2 \rangle,
\ee
where $v^{-\beta}\Psi(v) \to 0$, when $v \gg 1$. The large moments may be computed using the
saddle point approximation (see Appendix~\ref{appendix1}) to give
\begin{align}
\label{eq:asy1}
&\langle|v-v_1|^{\tilde{\delta}} v^{2n}\rangle \sim n^{\frac{1+\tilde{\delta}}{\beta}-\frac{1}{2}} \left[\frac{2 n}{a e \beta}\right]^{\frac{2 n}{\beta}}
e^{\Psi\left[\left(\frac{2n}{a \beta}\right)^{\frac{1}{\beta}}\right]},\\
&\langle|v-v_1|^{\delta} v^{2n-k}v_1^{k}\rangle \sim 
n^{\frac{2+\delta}{\beta}-1} \left[\frac{ n}{a e \beta}\right]^{\frac{2 n}{\beta}}
\times \label{eq:asy2} \\
&e^{n\left[ \frac{2-x}{\beta} \ln (2-x)+ \frac{x}{\beta} \ln (x)\right]}
e^{\Psi\left[\left(\frac{(2-x)n}{a \beta}\right)^{\frac{1}{\beta}}\right]+\Psi\left[\left(\frac{x n}{a \beta}\right)^{\frac{1}{\beta}}\right]},
\notag
\end{align}
where $n \gg 1$ and where $x=k/n$, $k,n \gg 1$.  A similar analysis shows that the moments of the noise distribution $\Phi(\eta)$  
[as in \eref{eq:noise-pdf}] has the 
asymptotic behavior
\be
N_{2n} \sim n^{\frac{1+\tilde{\chi}}{\gamma}-\frac{1}{2}} \left[\frac{2 n}{b e \gamma}\right]^{\frac{2 n}{\gamma}}, ~~ n \gg 1.
\label{eq:noise-asym}
\ee

The $(2n)^{th}$ moment of velocity diverges exponentially with $n$. Thus, each term in 
the right hand side of \eref{moment-main} is exponentially large, and the  sums
can be replaced by the largest term with negligible error. This largest term may belong
to either the first or second sum.  By keeping in \eref{moment-main} only the terms due to inter-particle collision, we obtain
\begin{align}
\label{moment-collision}
&
\langle|v-v_1|^{\tilde{\delta}} v^{2n}\rangle
\sim
\notag \\
&
\sum_{k=1}^{2 n-1} \binom{2n}{k}\alpha^{k}(1-\alpha)^{2n-k} \langle|v-v_1|^\delta  v^{2n-k}v_1^{k}\rangle, ~n \gg 1,
\end{align}
whose self-consistent solution for $\beta$ will be denoted by $\beta_c$. Likewise, we denote by $\beta_d$ the self-consistent solution of $\beta$ for the equation
obtained by dropping the collision term in \eref{moment-main}:
\begin{equation}
\label{moment-driving}
\langle|v-v_1|^{\tilde{\delta}} v^{2n}\rangle
\sim
\sum_{k=1}^{n}\binom{2n}{2k} r_w^{2n-2k} \langle v^{2n-2k}\rangle  N_{2 k}, ~n \gg 1.
\end{equation}
Clearly, the true solution for $\beta$ is
\be
\label{eq:betaequality}
\beta=\min(\beta_d,\beta_c).
\ee
We now determine $\beta_c$ and $\beta_d$.

\subsection{\label{sec:collisionanalysis} Analysis of the collision term}

We first analyze \eref{moment-collision}, that arises  from keeping terms due to inter-particle collisions, and show that the only possible solution  is 
$\beta_c=\infty$.
Let $k=x n$, where $ 0 \leq x \leq 2$. The summation on the right hand side of \eref{moment-collision} may now be converted into an
integral, $\sum_k \to n \int dx$.  Approximating the factorials by the Stirling's approximation, and on replacing the left hand side by using \eref{eq:asy1}, \eref{moment-collision} reduces to
\be
\label{eq:1}
\frac{n^{\frac{1+\tilde{\delta}}{\beta_c}}}{\sqrt{n}} \left[\frac{2 n}{a e \beta_c}\right]^{\frac{2 n}{\beta_c}}
\sim
\frac{n^{\frac{2+\delta+2 \chi}{\beta_c}}}{\sqrt{n}} \left[\frac{ n}{a e \beta_c}\right]^{\frac{2 n}{\beta_c}}
\int dx e^{n f(x)},
\ee
where we have also substituted the terms in the right hand side of \eref{moment-collision} with the 
asymptotic form for the moments in \eref{eq:asy2}. The function $f(x)$ is given by
\be
f(x) = \ln 4 + x \ln \frac{\alpha}{x^{\frac{\beta_c-1}{\beta_c}}}+ (2-x) \ln \frac{1-\alpha}{(2-x)^{\frac{\beta_c-1}{\beta_c}}}.
\label{function in analysis collision term}
\ee
Also, in \eref{eq:1}, we have dropped the correction terms involving $\Psi$, as they will not be relevant for the analysis
that follows for determining $\beta_c$.
For large $n$, the integral is dominated by the region around $x^*$, the value of $x$ for which $f(x)$ is maximized.
From $f'(x^*)=0$, we obtain
\be
x^* = \frac{2}{1+ \left(\frac{1-\alpha}{\alpha}\right)^{\frac{\beta_c}{\beta_c-1}}}.
\ee
For $\beta_c >1$, $x^*$ is a local maximum while for $\beta_c<1$, $x^*$ is a local minimum and $f(x)$ is maximized
near the endpoints $x=0$ and $x=2$.   When $\beta_c=1$, $f(x)$ is either  identically zero when $\alpha=1/2$
or linear in $x$ with positive slope $\ln(\frac{\alpha}{1-\alpha})$, such that $f(x)$ is maximum at $x=2$.

Consider first $\beta_c > 1$. Then, doing the saddle point integration about $x^*$, \eref{eq:1} reduces to
\be
\label{eq:2}
\frac{n^{\frac{1+\tilde{\delta}}{\beta_c}}}{\sqrt{n}} e^{\frac{2 n \ln 2}{\beta_c}}
\sim
\frac{n^{\frac{2+\tilde{\delta}+2 \chi}{\beta_c}}}{n} 
e^{n f(x^*)},
\ee
where
\be
\label{eq:3}
f(x^*) = \frac{2}{ \beta_c} \ln 2 + 2 \ln \alpha + \frac{2 (\beta_c-1)}{\beta_c} \ln \frac{2}{x^*}.
\ee
We now show that $f(x^*) < \frac{2}{\beta_c} \ln 2$.
Since $1/2 \leq \alpha <1$ implies that $(1-\alpha)/\alpha\leq1$, and since $\beta_c /(\beta_c-1) >1$ for $\beta_c >1$, this  implies that
$[(1-\alpha)/\alpha]^{\beta_c /(\beta_c-1)}\leq (1-\alpha)/\alpha$. It is then straightforward to show that $x^* \geq 2 \alpha$.
Substituting into \eref{eq:3}, and using $\alpha<1$, we obtain
\be
\label{comparing exp collision sum}
f(x^*) \leq\frac{2}{\beta_c} \ln (2\alpha)< \frac{2}{\beta_c} \ln 2.
\ee
Now, comparing the exponential terms in \eref{eq:2}, the left hand side is exponentially larger than the right hand side,
and the only possible solution is
\be
\label{eq:betac}
\beta_c= \infty.
\ee
Thus, if we assume that $\beta_c>1$, then it implies that $\beta_c=\infty$.

Consider now the second case when $\beta_c \leq 1$. Now $f(x)$ is maximised at the endpoints (except when $\alpha=1/2,~ \beta_c=1$), 
and we examine \eref{moment-collision}
for $k=2,~ 2n-2$. However, both these terms are exponentially smaller than the left hand side due to presence of the
terms $\alpha^{2 n}$ or $(1-\alpha)^{2n}$, and hence no solution is possible. When $\alpha=1/2$ and $\beta_c=1$, then $f(x^*) =0$.
From \eref{eq:2}, the left hand side is exponentially larger than the right hand side and hence this is not a valid non-trivial solution.

We, thus, conclude that $\beta_c=\infty$, as in
\eref{eq:betac}, is the only possible solution.

\subsection{\label{sec:drivinganalysis} Analysis of driving term}

We now determine $\beta_d$ by analyzing \eref{moment-driving}, that was obtained from \eref{moment-main} by dropping the collision term and
retaining the driving term. 
Let $k=x n$, where $ 0 \leq x \leq 1$. The summation on the right hand side of \eref{moment-driving} may now be converted into an
integral, $\sum_k \to n \int dx$. Approximating the factorials by Stirling's approximation, and on replacing the moments on the left and right side of \eref{moment-driving} by using \eref{eq:asy1}, we obtain
\begin{align}
&\frac{n^{\frac{1+\tilde{\delta}}{\beta_d}}}{\sqrt{n}} 
\left[\frac{2 n}{a e \beta_d}\right]^{\frac{2 n}{\beta_d}}
e^{\Psi\left[ \left( \frac{2 n}{a \beta_d}\right)^{1/\beta_d}\right]}
\sim 
\frac{n^{\frac{1+\tilde{\chi}}{\gamma}}}{n^{\frac{1}{2}-\frac{1}{\beta_d}}} 
\left[\frac{2 n}{e}\right]^{\frac{2 n}{\beta_d}} \times \nonumber \\
&
 \int    dx n^{2 n x (\frac{1}{\gamma}-\frac{1}{\beta_d})} e^{2 n g(x)} e^{\Psi\left[ \left( \frac{2 n (1-x)}{a \beta_d}\right)^{1/\beta_d}\right]},
 \label{eq:10}
\end{align}
where the function $g(x)$ is given by
\bea
g(x) &=& \frac{1-\gamma}{\gamma} x \ln x+ \frac{1-\beta_d}{\beta_d} (1-x) \ln(1-x)\nonumber \\
&-&\frac{1-x}{\beta_d} \ln (a \beta_d)-\frac{x}{\gamma}\ln (b \gamma)+(1-x) \ln r_w.
\label{eq:11}
\eea
We consider the three possible cases $\beta_d>\gamma$, $\beta_d<\gamma$, and $\beta_d=\gamma$ separately.

\subsubsection{Case 1: $\beta_d > \gamma$}

In this case, the integrand of \eref{eq:10} is dominated by the term $n^{2 n x (\frac{1}{\gamma}-\frac{1}{\beta_d})}$. Clearly, this term
is maximized when $x=1$. This corresponds to the term $k=n$ in \eref{moment-driving}, such that 
\be
\label{eq:12}
\frac{n^{\frac{1+\tilde{\delta}}{\beta_d}}}{\sqrt{n}} 
\left[\frac{2 n}{a e \beta_d}\right]^{\frac{2 n}{\beta_d}}
e^{\Psi\left[ \left( \frac{2 n}{a \beta_d}\right)^{1/\beta_d}\right]}
\sim
N_{2n}.
\ee
Comparing with \eref{eq:noise-asym}, we 
immediately obtain $\beta_d=\gamma$. However, this contradicts our assumption that $\beta_d > \gamma$, and hence no solution is
possible for this case.

\subsubsection{Case 2: $\beta_d < \gamma$}
\label{subsection betad lt gamma}
We now consider the case $\beta_d < \gamma$. If no valid solution is found for this case, then the only remaining solution
is $\beta_d=\gamma$. Thus, if a valid solution is found, then the actual solution for $\beta_d$ will be the smaller of  this value and $\gamma$.

When $\beta_d < \gamma$,  the term $n^{2 n x (\frac{1}{\gamma}-\frac{1}{\beta_d})}$ in the integrand of \eref{eq:10}
is maximized when $x=0$, which corresponds to the term $k/n \to 0$. However, this does not imply that the largest
term in the sum in \eref{moment-driving} corresponds to $k=1$. Rather, it could be at some $k^*$ that scales with
$n$ as $k^* \sim n^\zeta$, where $\zeta <1$. 

To do a more detailed analysis near $k=0$, we rewrite \eref{moment-driving}  as
\begin{equation}
\label{moment-driving1}
\langle|v-v_1|^{\tilde{\delta}} v^{2n}\rangle
\sim
\lambda_d \sum_{k=1}^{n}t_k, ~~n \gg 1.
\end{equation}
where
\begin{equation}
\label{tk}
t_k=\binom{2n}{2k} r_w^{2n-2k} \langle v^{2n-2k}\rangle  N_{2 k}.
\end{equation}

First consider the ratio between the first two terms, $t_2/t_1$, which after rearrangement is equal to
\be
\frac{t_2}{t_1}=\frac{n^{2-2/\beta_d}}{3r_w^2}\frac{N_4}{N_2} \left(\frac{a\beta_d}{2}\right)^\frac{2}{\beta_d}.
\label{eq:ratio1}
\ee
Clearly, if $\beta_d<1$, then $t_2 \ll t_1$ for large $n$. Thus, for $\beta_d<1$ it is possible to replace the sum on the right hand side of \eref{moment-driving1}
by the first term $t_1$, which results in:
\be
2\lambda_c\langle v^{2n+\tilde{\delta}}\rangle=\lambda_d\binom{2n}{2} r_w^{2n-2}\langle v^{2n-2}\rangle N_2,~~n\gg1.
\label{beta lt gamma moments compare}
\ee 
Using \eref{eq:asy1} to replace the moments of velocity in \eref{beta lt gamma moments compare}, we obtain:
\be
\left(\frac{2n}{a\beta_d}\right)^{\frac{2+\tilde{\delta}}{\beta_d}}=\frac{\lambda_c}{\lambda_d}n^2 r_w^{2n-2}N_2.
\label{beta lt gamma t1 compare}
\ee
When $r_w \neq 1$, the right hand side is exponentially smaller than the left hand side and \eref{beta lt gamma t1 compare}  does not have
a valid solution. However, when $r_w=1$, by comparing the powers of $n$ on both sides of 
\eref{beta lt gamma t1 compare}, we obtain $\beta_d=1+\delta/2$. Since, this analysis is valid only for $\beta_d<1$, we obtain the
constraint that $\delta<0$.  When $\delta=0$ and $r_w=1$, the velocity distribution may be determined exactly~\cite{Prasad:13,Prasad:17}.
For $\gamma \geq1$, $\beta_d=1$, and $a=\omega^*$, where $\omega^*$ is the solution for $1=\lambda_d f(\omega)$, with $\lambda_d$ being
the rate of driving and $f(\omega)=\langle e^{-i\omega\eta}\rangle_{\Phi}$ being the characteristic function of the noise distribution.
We can, thus, summarise these results:
\be
\beta_d = \min\left[\frac{2+\delta}{2},\gamma\right], ~~ \delta\leq 0, ~r_w=1,
\label{eq:betad1}
\ee
 where the minimum criterion arises from our assumption $\beta_d < \gamma$,
with the constant $a$ given by
\be
a=\begin{cases}\frac{4}{2+\delta} \sqrt{\frac{\lambda_d}{\lambda_c N_2}}, & \delta<0, ~r_w=1,~\frac{2+\delta}{2}<\gamma,\\
\omega^*, & \delta=0, ~r_w=1, ~\gamma \geq1.
\end{cases}
\label{eq:constanta}
\ee

Now consider the ratio in \eref{eq:ratio1} when $\beta_d >1$. From \eref{eq:ratio1}, it is clear that $t_2 \gg t_1$ for large $n$. To find $k^*$ for which
$t_{k^*}$ is the largest term, we consider the scaling  $k \sim n^\zeta$ with $\zeta<1$. Doing a 
change of variables $k= y n^\zeta$  and  converting the sum in \eref{moment-driving1} to an integral, we obtain
\begin{align}
\langle|v-v_1|^{\tilde{\delta}} v^{2n}\rangle
&\sim n^{\zeta\left[\frac{(1+\tilde{\chi})}{\gamma}-\frac{1}{2}\right]+\frac{1}{\beta_d}-\frac{1}{2}}\left(\frac{2n}{ae\beta_d}\right)^\frac{2n}{\beta_d}\times \nonumber \\
&e^{\Psi([\frac{2n}{a\beta_d}]^\frac{1}{\beta_d})}r_w^{2n}
\int dy e^{2n^\zeta h(y)},
\label{RHS  n tau expansion beta lt gamma}
\end{align}
where the function $h(y)$ is give by
\begin{align}
h(y)=&y \left(1-\frac{1}{\beta_d}-\zeta\left[1-\frac{1}{\gamma}\right]\right)\ln n-y\ln y+y-y\ln r_w\notag\\&-\frac{y}{\beta_d}\ln \left(\frac{2}{a\beta_d}\right)+\frac{y}{\gamma}\ln\left(\frac{2y}{be\gamma}\right).
\label{beta lt gamma n tau expansion}
\end{align}
Note that the function $h(y)$ has a term proportional to $\ln n$ in addition to an $n$-independent term. Since the term proportional
to $\ln n$ is also linearly proportional to $y$, $h(y)$ is maximized at $y=0$ or $y=\infty$ depending on the sign of the coefficient.
But both of these solutions imply that there is no nontrivial solution for $\zeta$. The only other option is that the coefficient
of $\ln n$ is identical to zero, enabling us to determine $\zeta$:
\be
\zeta=\frac{1-\beta_d^{-1}}{1-\gamma^{-1}}.
\label{tau}
\ee
For this value of $\zeta$, the function $h(y)$ is largest when
\be
y^*=\left(\frac{2}{b\gamma}\right)^{\frac{1}{\gamma-1}}\left[\left(\frac{2}{a\beta_d}\right)^\frac{1}{\beta_d}r_w\right]^{-\frac{\gamma}{\gamma-1}}.
\ee
Expanding about $y^*$ and doing a saddle point integration, \eref{RHS  n tau expansion beta lt gamma}  may be simplified to
\begin{align}
&\frac{n^{\frac{1+\tilde{\delta}}{\beta_d}}}{\sqrt{n}} 
\left[\frac{2 n}{a e \beta_d}\right]^{\frac{2 n}{\beta_d}}
e^{\Psi\left[ \left( \frac{2 n}{a \beta_d}\right)^{1/\beta_d}\right]}
\sim n^{\zeta\left[\frac{(1+\tilde{\chi})}{\gamma}-\frac{1}{2}\right]+\frac{1}{\beta_d}-\frac{1}{2}} \times \nonumber\\
&\left(\frac{2n}{ae\beta_d}\right)^\frac{2n}{\beta_d}
e^{\Psi\left[ \left( \frac{2 n}{a \beta_d}\right)^{1/\beta_d}\right]}
r_w^{2n} e^{2n^\zeta (\gamma-1) \gamma^{-1} y^*}
\label{eq:final1}
\end{align}
where we have replaced the left hand side of \eref{RHS  n tau expansion beta lt gamma} using \eref{eq:asy1}.

It is clear that no self-consistent solution for \eref{eq:final1} is possible unless $r_w=1$ and $\zeta=0$. Else, the right hand side is always
(due to $r_w^{2n}$ term and the last term) either exponentially smaller or larger than the left hand side.  However, from \eref{tau}, 
we  know that $\zeta>0$, since we have assumed that
$\beta_d >1$. Therefore, there is no self-consistent solution possible for \eref{eq:final1}.

This analysis shows that when $r_w<1$, a solution satisfying $\beta_d<\gamma$ is not possible. However, it
raises the intriguing possibility of such a solution when $r_w=1$ and $\beta_d=1$ (corresponding to $\zeta=0$). In this
case, we would expect that $k^*$ scales as some power of $\ln n$. Such scaling would, in turn, introduce logarithmic corrections
to the velocity distribution. In the remainder of this section, we show that such a self-consistent solution may indeed be
found for $\delta>0$, thus extending the result in \eref{eq:constanta} to any $\delta$.

To this end, let
\bea
P(v)\sim e^{-a|v| (\ln |v|)^\xi},~\delta>0, ~ r_w=1, ~ v^2  \gg \langle v^2 \rangle.
\label{eq:43}
\eea
The large moments of velocity may be determined by generalizing the calculation in the Appendix~\ref{appendix1} to give
\bea
\langle|v-v_1|^{\tilde{\delta}} v^{2n}\rangle 
\sim \left[\frac{n}{(\ln n)^\xi}\right]^{\tilde{\delta}} M_{2n},
\label{moments  pdf log correction0}
\eea
where
\be
M_{2n}\sim \frac{\sqrt{n}}{(\ln n)^{\xi}} \left[\frac{2 n}{a e (\ln n)^\xi}\right]^{2 n}
e^{\Psi \left[\frac{2n}{a (\ln n)^\xi}\right]} e^{2 n \xi^2\frac{\ln(\ln n)}{\ln n}}.
\label{moments  pdf log correction1}
\ee
Consider now the ratio of two consecutive 
$t_k$ [see \eref{tk}] with $r_w=1$. Substituting for the asymptotic forms from \eref{moments  pdf log correction1} and \eref{eq:noise-asym}, we
obtain in the limit $n,k \gg 1$, and $k/n \to 0$:
\be
\frac{t_{k+1}}{t_k} \approx \left[ 
\frac{a (\ln n)^\xi}{2 k} \left(\frac{2 k}{b \gamma}\right)^{\gamma^{-1}}
\right]^2.
\label{eq:tkratio10}
\ee
The stationary point $k^*$ satisfies $t_{k^*+1} \approx t_{k^*}$. Clearly $k^*$ has the form
\be
k^* = z^* (\ln n)^\theta.
\ee
Substituting into \eref{eq:tkratio10} and equating to $1$, we obtain
\bea
\theta&=& \frac{\xi\gamma}{\gamma-1},
\label{expression for zeta tau}  \\
z^* &=& \frac{1}{2} \left(\frac{a^\gamma}{b\gamma}\right)^{\frac{1}{\gamma-1}}.
\label{zstar}
\eea

The moment equation [\eref{moment-driving1}] with $t_k$, as in \eref{tk}, now reduces to
\be
\left[\frac{n}{(\ln n)^\xi}\right]^{\delta} M_{2n} \sim \binom{2n}{2 k^*} M_{2n - 2 k^*}  N_{2 k^*}
\label{eq:50}
\ee
For $k \ll n$, it is straightforward to obtain from \eref{moments  pdf log correction1} that
\be
M_{2n - 2 k} \sim M_{2n} \left[ \frac{a (\ln n)^\xi} {2 n}\right]^{2k}
\ee
Substituting into \eref{eq:50}, replacing the factorials by Stirling's approximation, and using the
asymptotic form \eref{eq:noise-asym} for $N_{2 k^*}$, we obtain
\be
\left[\frac{n}{(\ln n)^\xi}\right]^{\delta}  \sim 
e^{\frac{2 z^* (\gamma-1)}{\gamma} (\ln n)^\theta}
(k^*)^{\frac{\tilde{\chi}+1}{\gamma}-\frac{1}{2}}
\ee
Comparing the leading order term (powers of $n$), it is clear that $\theta$ has to be equal to one for
a power law to appear on the right hand side. Then, comparing the powers of $n$, we obtain
\bea
\theta &=& 1,\\
a&=&\left[\left(\frac{\delta\gamma}{\gamma-1}\right)^{\gamma-1}b\gamma \right]^\frac{1}{\gamma}.
\eea
From \eref{expression for zeta tau}, when $\theta=1$, we obtain
\be
\xi=\frac{\gamma-1}{\gamma}.
\label{eq:55}
\ee

To summarize, for $r_w=1$, we obtain for $\delta>0$, logarithmic corrections
to the leading behaviour, as described in \eref{eq:43}, with $\xi$ given by \eref{eq:55}. This is valid when 
$\gamma >1$. For $\gamma <1$, $\beta_d=\gamma$.

\subsubsection{Case 3: $\beta_d = \gamma$}

We now consider the third and final case $\beta_d = \gamma$. 
Then the equation for moments, as given in  \eref{eq:10},  simplifies to
\be
n^{\frac{\tilde{\delta}}{\gamma}}
a^{\frac{-2 n}{\gamma}}
e^{\Psi\left[ \left( \frac{2 n}{a \gamma}\right)^{\frac{1}{\gamma}}\right]}
\sim 
n^{\frac{1+\tilde{\chi}}{\gamma}}
 \int    dx e^{2 n g(x)} e^{\Psi\left[ \left( \frac{2 n (1-x)}{a \gamma}\right)^{\frac{1}{\gamma}}\right]},
 \label{eq:15}
\ee
where the function $g(x)$ is now given by
\bea
g(x) &=& \frac{1-\gamma}{\gamma} \left[x \ln x+ (1-x) \ln(1-x)\right]\nonumber \\
&-&\frac{1-x}{\gamma} \ln a-\frac{x}{\gamma}\ln b+(1-x) \ln r_w.
\label{eq:16}
\eea
For large $n$, the main contribution to the integral in \eref{eq:15} is from the region around $x^*$, the value
of $x$ at which $g(x)$ is maximum. The solution to $g'(x^*)=0$ is
\be
x^*=\left[1+\left(\frac{b r_w^\gamma}{a}\right)^\frac{1}{\gamma-1}\right]^{-1},
\label{alpha1 maximum}
\ee
where
\be
g(x^*) = \frac{\gamma-1}{\gamma}\ln\left[ 1+\left(\frac{b  r_w^\gamma}{a}\right)^\frac{1}{\gamma-1} \right]-\frac{1}{\gamma} \ln(b).
\label{extremal f}
\ee
The second derivative $g''(x^*) = (1-\gamma)/\gamma [x^* (1-x^*)]^{-1}$ is negative
only for $\gamma >1$. 
 $x^*$ is a local minimum when $\gamma <1$, while for
$\gamma=1$, $g(x)$ is linear in $x$.  Thus, for $\gamma \leq 1$,  $g(x)$
takes on its maximum value at the endpoints $x=0$ or $x=1$.

First, consider the case when $\gamma \leq1$. Then the maximum corresponds to 
term $k=1$ or $k=n$ in \eref{moment-driving}. To determine which of these is the
larger one, we take the ratio of the $k=n$ term to the $k=1$ term. This ratio is
proportional to $r_w^{-2n}n^{2/\gamma -2}$ showing that the $k=n$ term is the largest.
Keeping only the $k=n$ term, \eref{moment-driving} reduces to
\be
\label{eq:17}
\frac{n^{\frac{1+\tilde{\delta}}{\gamma}}}{\sqrt{n}} 
\left[\frac{2 n}{a e \gamma}\right]^{\frac{2 n}{\gamma}}
e^{\Psi\left[ \left( \frac{2 n}{a \gamma}\right)^{1/\gamma}\right]}
\sim
N_{2n}
\sim n^{\frac{1+\tilde{\chi}}{\gamma}-\frac{1}{2}} \left[\frac{2 n}{b e \gamma}\right]^{\frac{2 n}{\gamma}}
\ee
where the asymptotic form for $N_{2n}$ has been obtained from  \eref{eq:noise-asym}.  
A solution is possible only when the correction term $\Psi(x) = \chi \ln |x|$. 
We then immediately obtain
\bea
\beta_d&=&\gamma, \nonumber \\
a&=&b,\qquad \gamma \leq 1, \label{eq:18}\\
\chi &=& \tilde{\chi}-\tilde{\delta}. \nonumber
\eea
 This solution is valid provided there exists no solution satisfying $\beta_d<\gamma$, as for $r_w=1$~[see~\sref{subsection betad lt gamma}].

Now, consider the case $\gamma >1$. Then $x^*$ maximizes $g(x)$ and 
\eref{eq:15} simplifies to
\be
n^{\frac{\tilde{\delta}}{\gamma}}
a^{\frac{-2 n}{\gamma}}
e^{\Psi\left[ \left( \frac{2 n}{a \gamma}\right)^{\frac{1}{\gamma}}\right]}
\sim 
\frac{n^{\frac{1+\tilde{\chi}}{\gamma}}}{\sqrt{n}}
e^{2 n g(x^*)+\Psi\left[ \left( \frac{2 n (1-x^*)}{a \gamma}\right)^{\frac{1}{\gamma}}\right]},
 \label{eq:19}
\ee
The coefficient $a$ as well as the correction term $\Psi$ may now be determined.
Equating the terms exponential in $n$, we obtain
\be
\frac{-1}{\gamma} \ln a=g(x^*).
\label{leading order exponential gamma gt1}
\ee
Substituting from \eref{extremal f} and solving for $a$, we obtain
\be
a=b\left(1-r_w^{\frac{\gamma}{\gamma-1}}\right)^{\gamma-1}, \quad \gamma >1.
\label{equation for a gamma gt1}
\ee
Note that when $\gamma=2$, \eref{equation for a gamma gt1} reduces to $a=b(1-r_w^2)$ as obtained in Refs.~\cite{Prasad:13,Prasad:17}.

To determine the correction term $\Psi$, we proceed as follows. For $a$ as given in \eref{equation for a gamma gt1}, \eref{eq:19} simplifies to
\be
n^{\frac{\tilde{\delta}}{\gamma}}
e^{\Psi\left[ \left( \frac{2 n}{a \gamma}\right)^{\frac{1}{\gamma}}\right]}
\sim 
\frac{n^{\frac{1+\tilde{\chi}}{\gamma}}}{\sqrt{n}}
e^{\Psi\left[ \left( \frac{2 n (1-x^*)}{a \gamma}\right)^{\frac{1}{\gamma}}\right]},
 \label{eq:20}
\ee
First, we assume that $\Psi(y) \sim y^{\beta'}$ with $\beta'<\beta_d$. However, it is straightforward to check
that \eref{eq:20} is not satisfied for this form of $\Psi$ since $x^* \neq 0$. 

Consider now $\Psi(y) = -a' (\ln y)^{\tau}$. For large $n$, \eref{eq:20} may be written as
\begin{align}
&n^{\frac{\tilde{\delta}}{\gamma}}
e^{-a' \gamma^{-\tau} (\ln n)^\tau  - a' \tau \gamma^{-\tau} (\ln n)^{\tau-1} \ln\frac{2}{a \gamma} }
\sim  \nonumber\\
&\frac{n^{\frac{1+\tilde{\chi}}{\gamma}}}{\sqrt{n}}
e^{-a' \gamma^{-\tau} (\ln n)^\tau  -a' \tau \gamma^{-\tau} (\ln n)^{\tau-1} \ln\frac{2(1-x^*)}{a \gamma} }.
 \label{eq:21}
\end{align}
The terms proportional to $e^{a' \gamma^{-\tau} (\ln n)^\tau}$ are equal on both sides of \eref{eq:21}. However,
the leading subleading terms can be matched only if 
\be
\tau=2, ~\quad \gamma >1.
\label{tau equation}
\ee
Equating the subleading terms with the power law corrections, we immediately obtain
\be
\label{eq:aprime}
a'=\frac{\gamma(\gamma-1)}{2\ln(r_w)}\left(  \frac{1}{\gamma}+\frac{\tilde\chi}{\gamma}- \frac{\tilde{\delta}}{\gamma}
-\frac{1}{2} \right), \quad \gamma >1.
\ee

The \eref{tau equation} suggests that  the tail for $\gamma>1$ does not allow a  solution with logarithmic correction of linear order i.e. $O(\ln n)$, or equivalently a power-law correction which results in
 \bea
 \chi=0,~~~~\gamma>1.
 \label{chi relation for gamma gt 1}
  \eea
Further from \eref{eq:aprime}, one may notice that $a^\prime$ does not vanish even when $\tilde\chi=0$. This indicates that the presence of the sub-leading correction in the velocity distribution is not a consequence of any subleading behaviour in the noise distribution.

\subsection{\label{sec:solution} Solution for $\beta$}

The solution for $\beta$ may now be found from \eref{eq:betaequality}.
Since $\beta_c=\infty$ [see \eref{eq:betac}], the velocity distribution depends only
on the driving term and $\beta=\beta_d$. The result for $\beta$ may be summarized as follows.
When $r_w<1$, $\beta=\gamma$. When $\gamma<1$ the parameters $a$ and $\chi$ may be expressed in terms
of the noise parameters as in \eref{eq:18}. When $\gamma>1$, there are additional logarithmic corrections as
in $\Psi(y) = a' (\ln y)^2$, where the parameters $a, a'$ are expressed in terms of the noise parameters as in
\eref{equation for a gamma gt1} and \eref{eq:aprime}. When $r_w=1$, we obtained that $\beta= \min[(2+\delta)/2,\gamma]$ for
$\delta\leq 0$
[see \eref{eq:betad1}]. For $\delta>0$ and $r_w=1$, the velocity distribution has logarithmic corrections
as described in \eref{eq:43} with $\xi$ as given in \eref{eq:55}. We numerically confirm some of these results in the Sec.~\ref{sec:numerical}.

\section{\label{sec:numerical} Numerical solution for $\delta=0$}

In this section, we study the moment equations numerically and confirm some of the analytically obtained
results. We limit our analysis to the case $\delta=0$, corresponding to Maxwell gas, when the moment equations
have a simpler form that is amenable to numerical analysis. We follow the same procedure as that used in
Ref.~\cite{Prasad:17}, where $\beta$ was determined numerically. We briefly summarize the procedure 
below.

\begin{figure}
\begin{center}
\includegraphics[width=0.6\hsize]{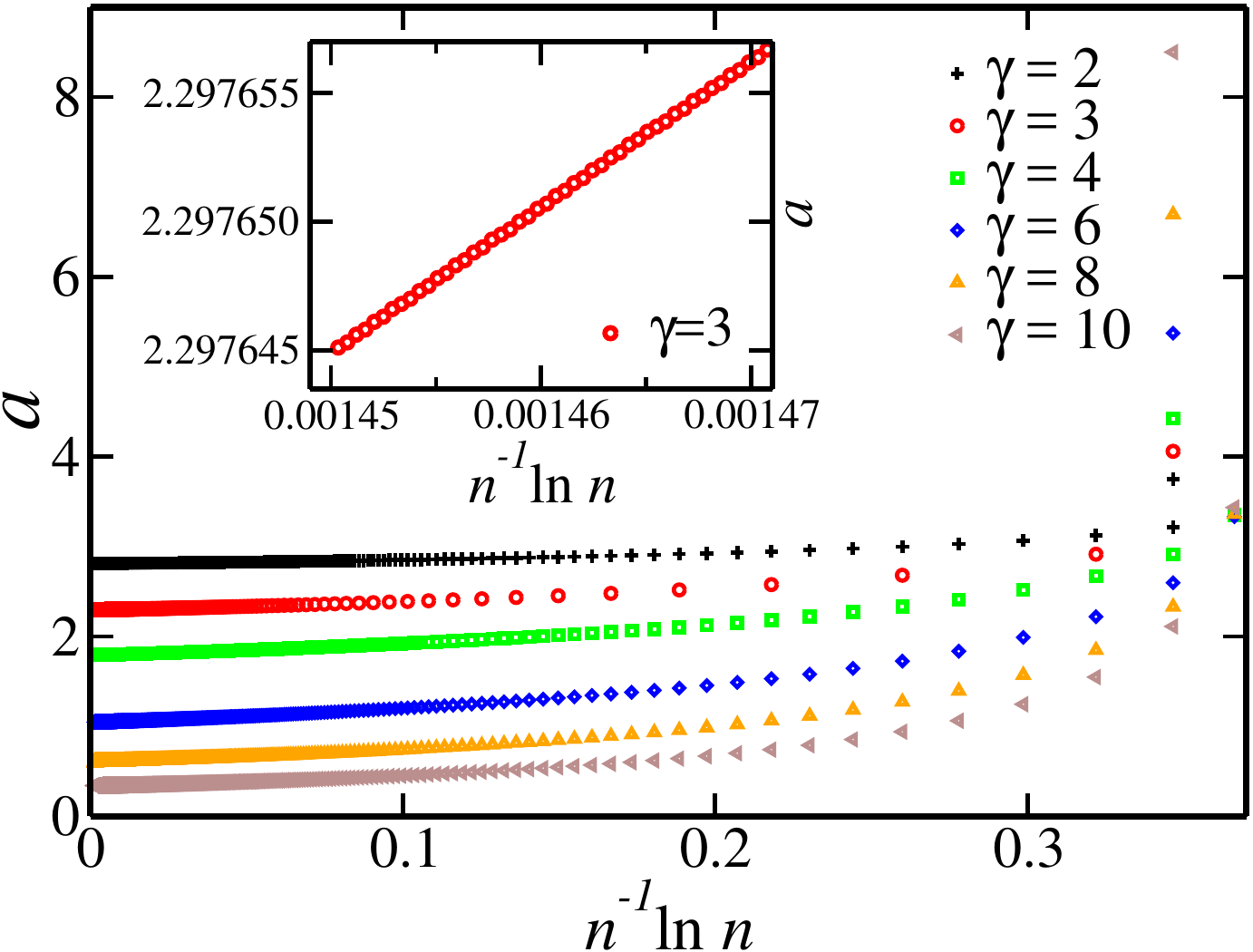}
\caption{(color online) The variation of $a(n)$, as defined in \eref{numerical a}, with  $n^{-1} \ln n$
for stretched exponential noise  distribution, as defined in  \eref{ansatz noise pdf}, with $b=3$ and for different values of $\gamma$.
The inset shows the same quantity, $a(n)$ in the limit of large $n$ (small values of  $n^{-1}\ln n$) for $\gamma=3$ to illustrate the linear dependence
for large  $n$.}
\label{Fig1}
\end{center}
\end{figure}
\begin{figure}
\begin{center}
\includegraphics[width=0.6\hsize]{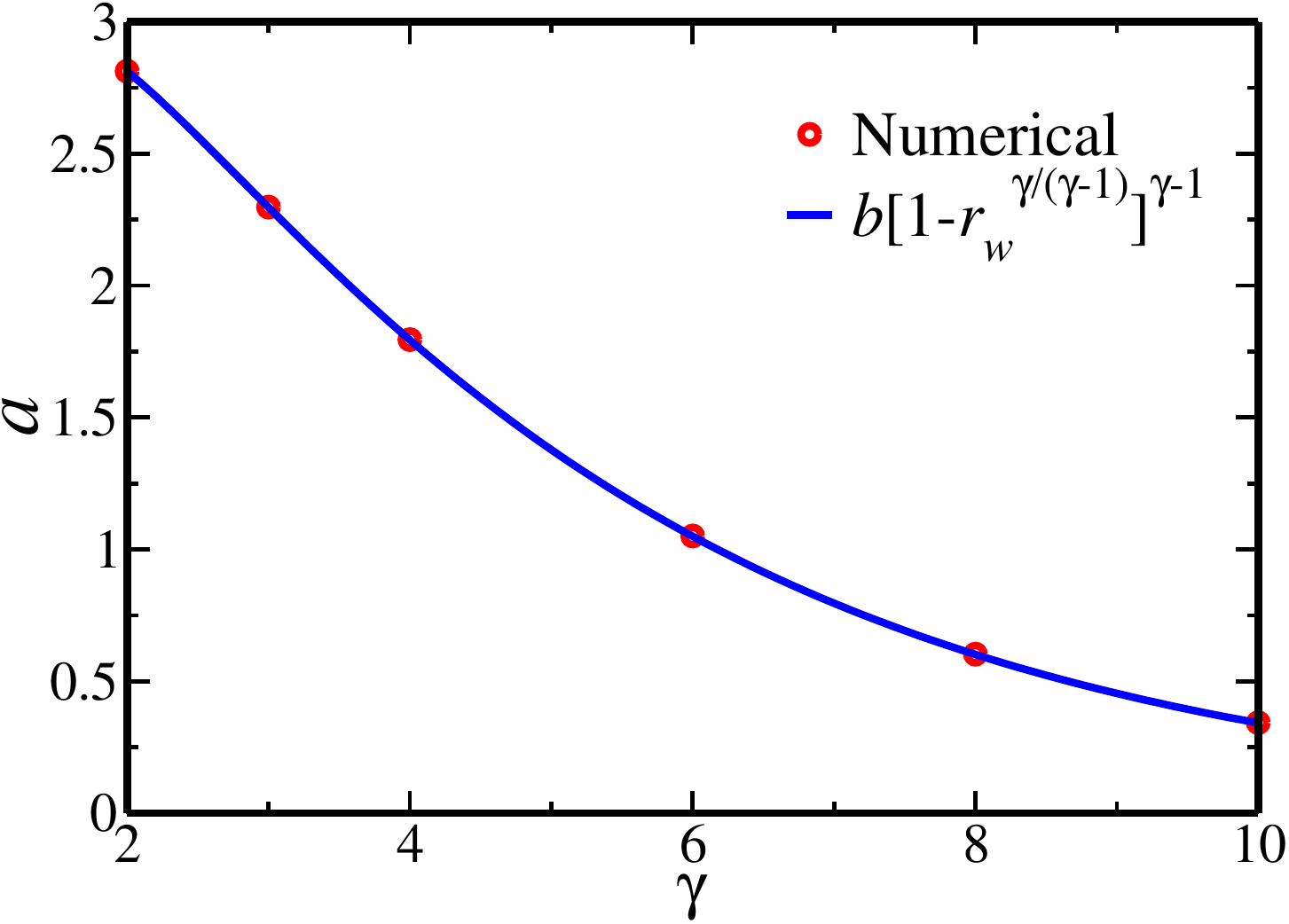}
\caption{(color online) Comparison of numerical value of the constant $a$, obtained by extrapolating $a(n)$ to $n \to \infty$ [see
\fref{Fig1}], with the analytical formula  [see \eref{equation for a gamma gt1}]. The data are for stretched exponential noise  distribution, as defined in  \eref{ansatz noise pdf}, with $b=3$ and for different values of $\gamma$.}
\label{Fig2}
\end{center}
\end{figure}

When $\delta=0$, the moment equation \eref{moment ratio delta non zero} simplifies to
\begin{align}
\label{moment ratio}
&\left[1-\alpha^{2n} - \left(1-\alpha \right)^{2n} +\frac{\lambda_d}{2 \lambda_c} \left(1-r_w^{2n}\right)\right]
M_{2n}=
\notag \\
&\sum_{k=1}^{n-1}\binom{2n}{2k}\alpha^{2k}(1-\alpha)^{2n-2k}M_{2k}M_{2n-2k}
\notag
\\
&+\frac{\lambda_d}{2 \lambda_c} \sum_{k=1}^{n}\binom{2n}{2k} r_w^{2n-2k} M_{2n-2k}
N_{2k}, 
\end{align} 
where $N_{i}$ is the $i$-th moment of the noise distribution, and
\be
M_{2n}\equiv\langle v^{2n}\rangle, \quad n=1,2,\ldots.
\ee
This relation expresses a moment in terms of lower order moments. Thus, by knowing
$M_0=1$, higher order moments may be obtained by iteration.

We will consider normalized stretched exponential distributions for the noise distribution $\Phi(\eta)$:
\be
\Phi(\eta)=\frac{b^{\frac{1}{\gamma}}}{2\Gamma\left(1+\frac{1}{\gamma}\right)}\exp(-b|\eta|^\gamma),~~~
b,\gamma>0.
\label{ansatz noise pdf}
\ee
for which the moments are
\bea
N_{2n}=b^{-2 n/\gamma} \frac{\Gamma(\frac{2n+1}{\gamma})}{\gamma \Gamma(1+\frac{1}{\gamma})}.
\eea
Consider now the velocity distribution as derived in this paper [see \eref{velocity pdf with log correction 1}]
$P(v) \sim \exp(-a|v|^\gamma-a' (\ln|v|)^2)$. For this distribution, from \eref{eq:asy1} with $\delta=0$, it is straightforward to show
that the ratio of consecutive moments has the form
\begin{equation}
\Delta_n \equiv \frac{M_{2n+2}}{M_{2n}}=\left(\frac{2n}{a\gamma}\right)^\frac{2}{\gamma}
\left[1-\frac{2 a'}{\gamma^2}\frac{\ln n}{n} +\mathcal{O}\left( \frac{1}{n}\right)\right]. 
\label{moment ratio with log correction leading order}
\end{equation}

By measuring $\Delta_n$ for large $n$, it was shown in Ref.~\cite{Prasad:17} that $\beta=\gamma$. Here, we focus on determining
$a$ and $a_1$ numerically and comparing with  \eref{velocity pdf with log correction 1}. In \fref{Fig1}, we show the variation of
the numerically computed
\be
a(n) = \frac{2n}{\gamma \Delta_n^{\gamma/2}},
\label{numerical a}
\ee
with $\ln n /n$. The data for each $\gamma > 1$ lie on a straight line which when extrapolated to infinite $n$ gives the numerical
estimate for $a$. \Fref{Fig2} compares the numerically obtained value with the analytically obtained expression, and we find an excellent agreement.

\begin{figure}
\begin{center}
\includegraphics[width=0.6\hsize]{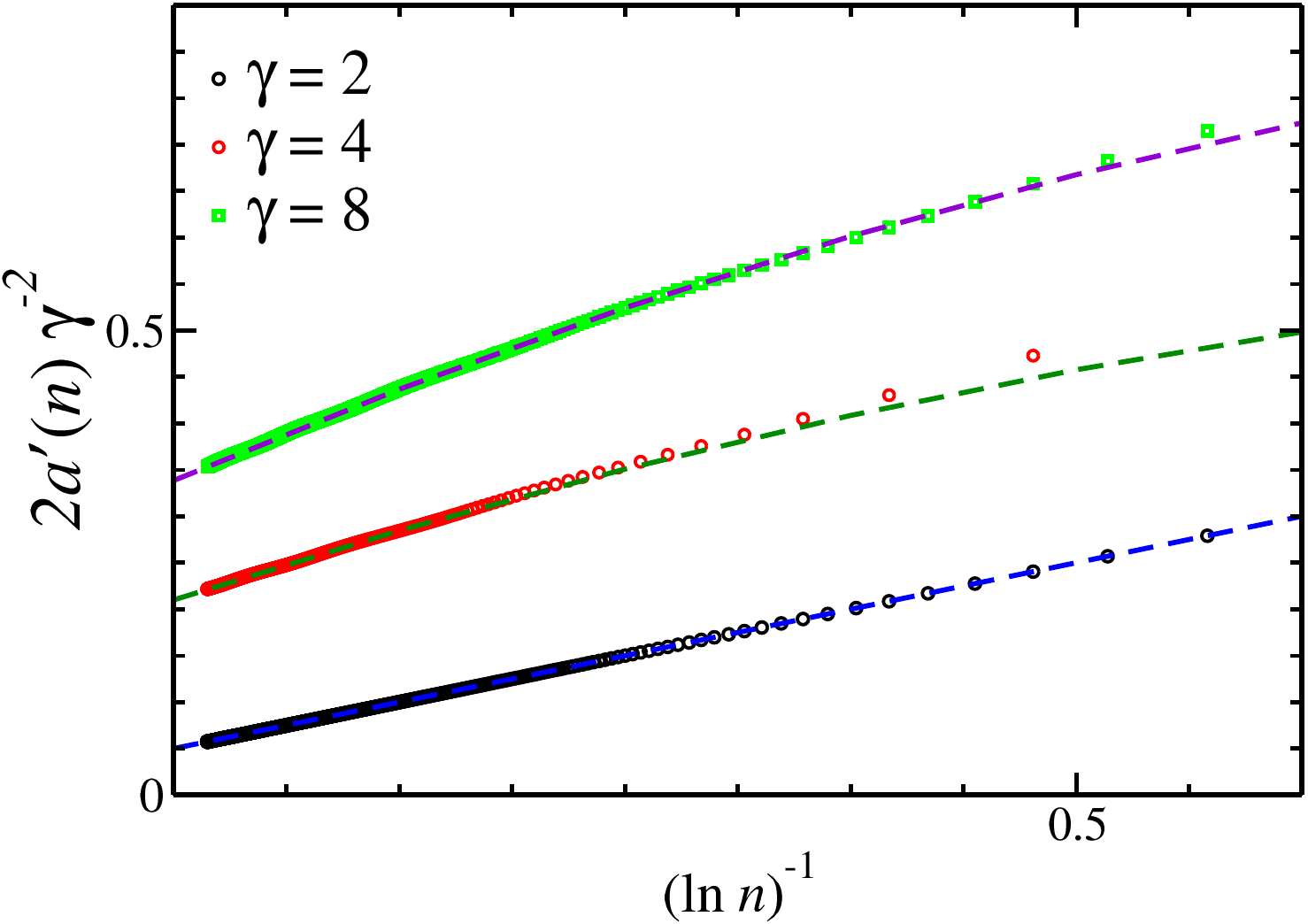}
\caption{(color online) 
The variation of numerical value of $2a^\prime(n)/\gamma^2$, as defined in \eref{a prime moment ratio with log correction},  with $[\ln n]^{-1}$
for stretched exponential noise  distribution, as defined in  \eref{ansatz noise pdf}, with $b=3$ and for different values of $\gamma$. 
The quadratic fits to the data are shown by dashed lines.}
\label{Fig3}
\end{center}
\end{figure}

\begin{figure}
\begin{center}
\includegraphics[width=0.6\hsize]{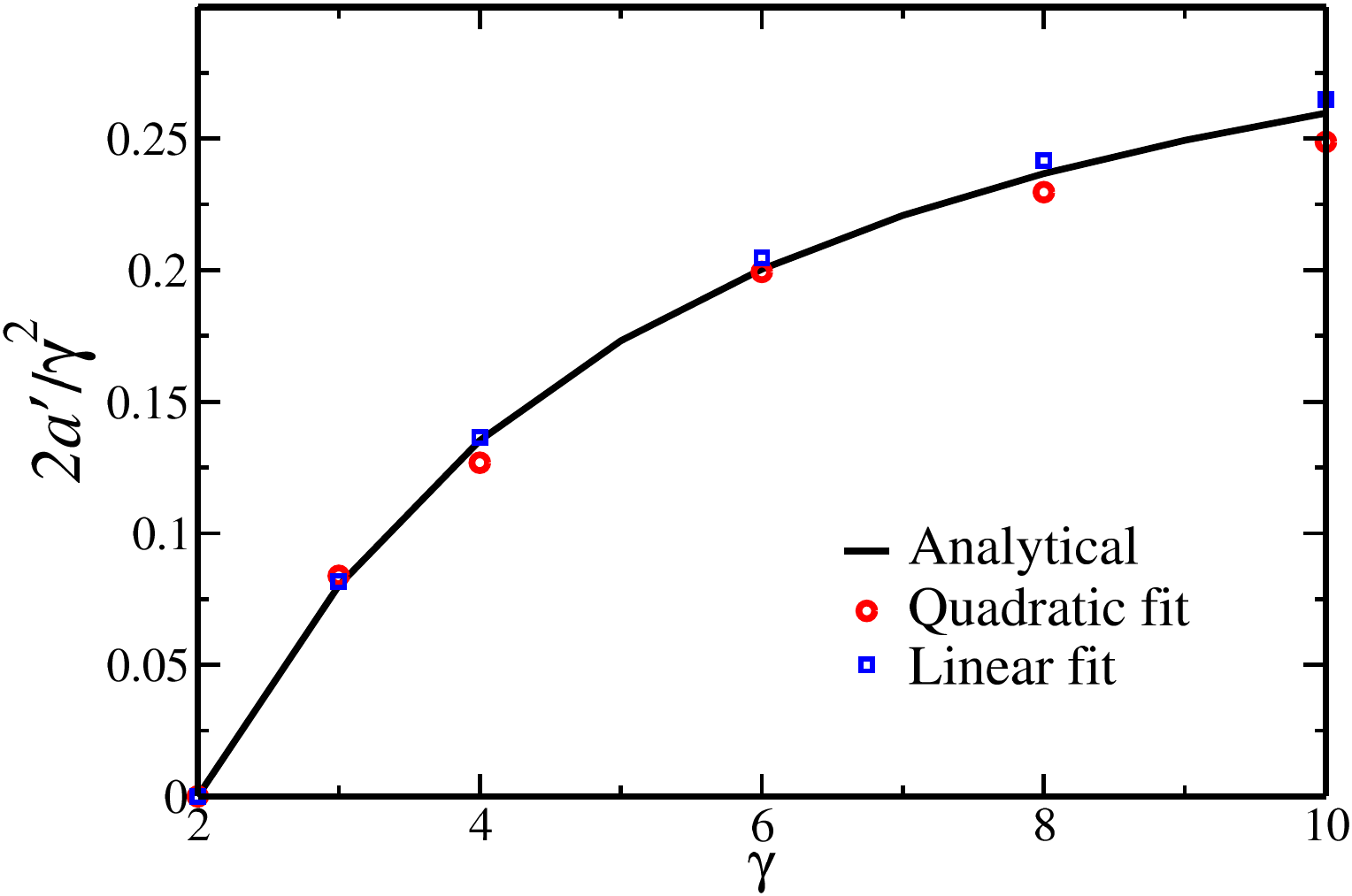}
\caption{(color online)  Comparison of numerical value of the constant $2a^\prime/\gamma^2$,  
obtained by extrapolating $2a^\prime(n)/\gamma^2$ to $n \to \infty$ [see
\fref{Fig3}], with the analytical formula [see \eref{eq:aprime} with $\tilde{\chi}=0,~\tilde{\delta}=0$] for two
different extrapolation schemes. The data are for stretched exponential noise  distribution, as defined in  \eref{ansatz noise pdf}, with $b=3$ and for different values of $\gamma$.  
}
\label{Fig4}
\end{center}
\end{figure}

Now, we determine the constant $a'$ by assuming that the expression for $a$ is exact. In that case, we define
\begin{equation}
\frac{2a'(n)}{\gamma^2} = \frac{n}{\ln n}\left[1-\Delta_n \left(\frac{a \gamma}{2 n}\right)^\frac{2}{\gamma}\right]
\label{a prime moment ratio with log correction}
\end{equation}
such that $a' = a'(n) + \mathcal{O}(1/\ln n)$.
The variation of $a'(n)$ with $1/\ln n$ is shown in \fref{Fig3}. For large $n$, the variation is linear. By extrapolating
to infinite $n$, we obtain $a'$. The extrapolation may be done using a linear or quadratic fit. The results for
both extrapolations are shown in \fref{Fig4} and compared with the analytically obtained result~[\eref{eq:aprime}] when $\tilde{\xi}=0,
~\tilde{\delta}=0$]. There is
excellent agreement. 

\section{\label{sec:summary}Summary and conclusions}

In summary, we derived analytical results for the asymptotic behavior of the velocity distribution for a driven inelastic granular gas with 
one-dimensional scalar velocity. Each pair of particles  undergo inelastic collisions as described in \eref{interparticle collision},
at a rate that depends on
the relative velocity $\Delta v$ as $ |\Delta v|^\delta$. Each particle is driven at a constant rate by dissipative driving as described in \eref{wall collision}, and is characterized by the dissipation parameter $r_w$. To perform our analysis of the moment equation, we assume that the leading asymptotic  behaviour of the velocity distribution is a stretched exponential. We then obtain self-consistent solutions that are consistent with this assumption. We obtain analytical results for arbitrary coefficients of
restitution, arbitrary $\delta$, as well as generic noise distributions characterized by a stretched exponential exponent $\gamma$ 
[see \eref{eq:noise-pdf}].
The main results obtained in this paper are summarized in 
Eqs.~(\ref{velocity pdf with log correction 1})--(\ref{eq:last}). 
For the dissipative case, $r_w <1$, the statistics of the velocity distribution are non-universal in that the stretched
exponential exponent $\beta$ is determined solely by the noise distribution.  
For $r_w=1$, the distribution
is universal, provided $\gamma > \min[\frac{2+\delta}{2},1]$. In addition to the stretched exponential exponent $\beta$, we also determined the different coefficients that 
describe the asymptotic behavior of the velocity distribution. Surprisingly, we find the presence of  logarithmic corrections to the leading stretched exponential behavior. These corrections could be confirmed through a numerical solution of the exact moment equations for the special case $\delta=0$. We 
note that there are no exact results for models where $\delta \neq 0$.

\begin{figure}
\begin{center}
\includegraphics[width=0.6\columnwidth]{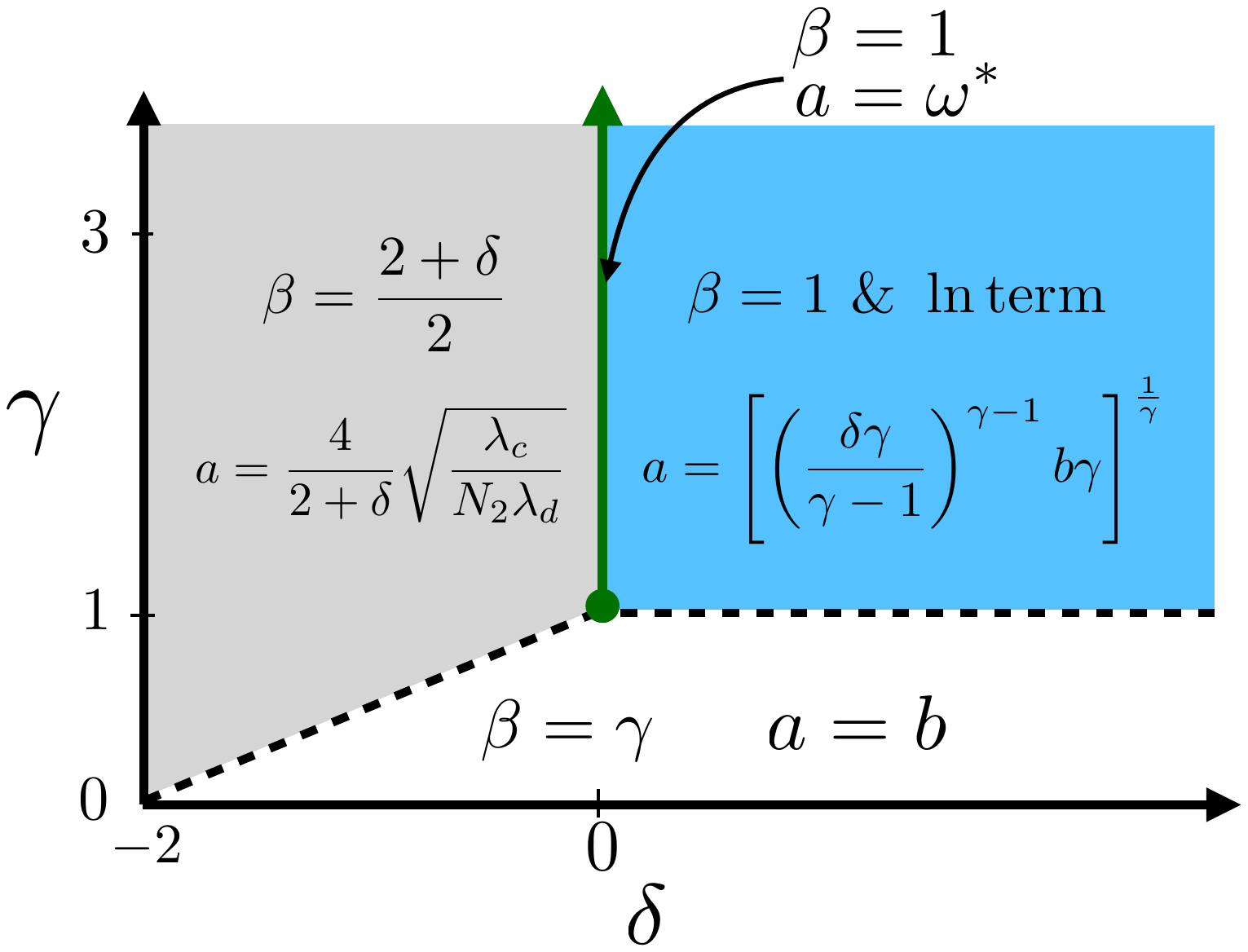}
\caption{(color online)  Summary of results for the stretched exponential exponent $\beta$ and the parameter $a$ as a function of
the parameters $\gamma$ and $\delta$,  for the case $r_w=1$.   
}
\label{Fig5}
\end{center}
\end{figure}

In the case of $r_w=1$ the solution for the stretched exponential exponent $\beta$ is equal to $(2+\delta)/2$, provided 
$\delta \leq 0$ and is equal to $1$ for $\delta>0$ albeit with additional logarithmic corrections. The results for $r_w=1$ depend sensitively
on the values of $\gamma$ and $\delta$ and are summarized in Fig.~\ref{Fig5}. The expression for $\beta=(2+\delta)/2$, when $-2<\delta\leq 0$,
is identical to that obtained in kinetic theory. However, while this result continues to also hold for positive $\delta$ in kinetic theory, our calculations show
that $\beta=1$ for $\delta>0$ with logarithmic corrections. Thus, the kinetic theory result of $\beta=3/2$ for $\delta=1$ (ballistic motion) cannot be obtained
from the solution presented in the current paper. However, if we had incorrectly truncated the expression for moments of noise in 
Eq.~(\ref{moment-driving1}) to
only the first term, then the kinetic theory result would be recovered. In terms of the master equation for the velocity distribution, the
difference between kinetic theory result and our results can be traced to  the
diffusive driving term in kinetic theory, that is obtained by truncating the Taylor expansion of the last term in \eref{eq:evolution}, in terms of
derivatives of the velocity distribution $P(v)$, up to only the second derivative.
However, it may be shown that this truncation is not valid when $\delta>0$, and maximal term in the Taylor expansion is a term corresponding to a 
higher order derivative [see Appendix~\ref{appendix2} for details of this analysis].

There is, however, a special limit in which the kinetic theory results may be recovered. Consider Eqs.~(\ref{moment-driving1}), \eqref{tk}, and 
\eqref{eq:ratio1} describing the relation between moments of velocity and moments of noise. Suppose in Eq.~(\ref{eq:ratio1}), the limit $N_2 \to 0$,
where $N_2$ is the second moment of the noise distribution, is taken before the limit $n \to \infty$ is taken. Then $N_4/N_2 \to 0$, and Eq.~(\ref{moment-driving1})
may be truncated by keeping only the first term on the right hand side. For a sensible limit we need to take $\lambda_d \to \infty$ keeping
$\lambda_d N_2$ fixed. For this special limit, we then obtain, for $r_w=1$, the kinetic theory result $\beta=(2+\delta)/2$.

This special limit was also considered in Ref.~\cite{Prasad:14}, where the driving mechanism (see \eref{wall collision}), in the  limit of diffusive driving, 
was shown to be an Ornstein-Uhlenbeck when the noise distribution is a gaussian~\cite{Prasad:14}.  In particular, the limit taken was 
$r_w\to \pm 1$ such that $(1\pm r_w) \lambda_d \to\Gamma$, and $N_2 \to 0$, $\lambda_d \to \infty$ keeping $\lambda_d N_2$ fixed. We note that in
Ref.~\cite{Prasad:14}, only the case $r_w \to -1$ was considered, but it is straightforward to see that the derivation goes through for $r_w \to 1$ also. For non-gaussian noise, as considered in the current paper, the derivation still goes through because in this limit, central limit theorem ensures that the
noise arising from repeated driving ($\lambda_d \to \infty$) is a gaussian. On including the dominant term arising from collisions, the resultant master equation for the 
steady state velocity distribution has the form
\be
-c_1 v^{\delta} P(v) + c_2 \frac{\partial^2}{\partial v^2} P(v)+ \Gamma \frac{\partial}{\partial v}[v P(v)] = 0,
\label{dissi_drive}
\ee
where $c_1$ and $c_2$ are constants. It is straightforward to see that, by balancing the second and third terms and ignoring
inter-particle collisions, one obtains $\beta=2$~\cite{Montanero:00,Biben02,Villani:06}. On the other hand, when $r_w=1$, then $\Gamma=0$, and on balancing
the first and third terms, one obtains $\beta=(2+\delta)/2$ as in kinetic theory~\cite{Ernst2003}.

 The analysis of the moment equation was performed under the assumption that  the leading asymptotic  behaviour of the velocity distribution is a stretched exponential. We then obtain self-consistent solutions that are consistent with this assumption. Thus, we are unable to show that the solutions that we obtain are unique, as there may be other self-consistent solutions with a different ansatz for the velocity distribution. However, the results that we obtain are in excellent agreement with the numerical results for the case $\delta=0$. Thus, we believe that the  results obtained in the paper may well be exact.

The results for the one dimensional driven granular gas obtained in this paper suggest that the velocity distribution is determined mostly
by the noise distribution and hence the hope for a universal distribution is misplaced. However, the one-dimensional gas is different from
a granular gas in higher dimensions as now the possibility of collisions which are not just head on are allowed.  In a recent paper, we have shown that, by  extending the above analysis to  a two-dimensional system,  the velocity distribution becomes universal with $\beta=2$ albeit with logarithmic corrections~\cite{Prasad:19}.

\appendix

\section{\label{appendix1}Large moments of the velocity distribution}

We determine the behavior of large moments when the velocity distribution has the asymptotic form:
\begin{equation}
P(v) \sim e^{-a |v|^\beta+\Psi(v)},~~|v|\to\infty,~a>0.
\label{eq appdx 1}
\end{equation}
where $v^{-\beta} \Psi(|v|)\to 0$ for $|v|\to\infty$.
The aim is to determine the moment
$\langle|v_1-v_2|^\delta v_1^{2n-k}v_2^k\rangle$ for large $n$ and $k$. In the limit $k \propto n$,
we change variables to  $k=xn$ with $0\le x\le1$, to obtain
\bea
\langle|v_1-v_2|^\delta v_1^{2n-k}v_2^{k}\rangle\sim\int dv_1dv_2|v_1-v_2|^\delta 
v_1^{(2-x)n}v_2^{xn}\notag\\\times e^{-a|v_1|^\beta-a|v_2|^\beta+\Psi(v_1)+\Psi(v_2)}\hspace{0.8cm}
\label{eq appdx 2}
\eea
      
Rescaling the integration variables $v_1$, and $v_2$ as
$v_1=n^{1/\beta}t_1$ and $v_2=n^{1/\beta}t_2$, \eref{eq appdx 2} simplifies to 
\begin{align}
\langle|v_1-v_2|^\delta v_1^{2n-k}v_2^{k}\rangle&\sim
n^{\frac{2+\delta + 2 n}{\beta}} \int dt_1dt_2e^{n[f(t_1)+ g(t_2)]}\times\notag\\
&|t_1-t_2|^\delta e^{\Psi(n^{1/\beta}t_1)+\Psi(n^{1/\beta}t_2)}
\label{eq appdx 3}
\end{align}
where,
\begin{align}
\begin{split}
f(t_1)&=-at_1^\beta+(2-x)\ln t_1,\\
g(t_2)&=-at_2^\beta+x\ln t_2.
\label{eq appdx 4}
\end{split}
\end{align}
The integral in \eref{eq appdx 3} may be evaluated using the 
saddle point approximation by maximizing $f(t_1)$ and $g(t_2)$ with respect to $t_1$ and $t_2$. Setting
$df(t_1)/dt_1=0$ and $dg(t_2)/dt_2=0$, we obtain the solution $t_1^*$ and $t_2^*$ to be
\begin{align}
t_1^*&=\left(\frac{2-x}{a\beta}\right)^\frac{1}{\beta},\\
t_2^*&=\left(\frac{x}{a\beta}\right)^\frac{1}{\beta},
\label{eq appdx 7}
\end{align}
and
\begin{align}
\begin{split}
f(t_1^*)&=\frac{2-x}{\beta}\ln\left(\frac{2-x}{ae\beta}\right),\\
g(t_2^*)&=\frac{x}{\beta}\ln\left(\frac{x}{ae\beta}\right).
\label{eq appdx 6}
\end{split}
\end{align}
Doing the saddle point integration in \eref{eq appdx 3}, we obtain
\begin{align}
&\langle|v_1-v_2|^\delta v_1^{2n-k}v_2^k\rangle\sim 
\frac{n^\frac{2+\delta}{\beta}}{n}\left(\frac{n}{2ae\beta}\right)^\frac{2n}{\beta}\times\hspace{1cm}\notag\\
&e^{n\left[\frac{2-x}{\beta}\ln(2-x)+
\frac{x}{\beta}\ln x\right]} e^{\Psi(n^{1/\beta}t_1^*)+\Psi(n^{1/\beta}t_2^*)},
\label{eq appdx 5}
\end{align}
where the extra factor of $n$  appears in the denominator of the left hand side of \eref{eq appdx 5} 
because of the two-dimensional saddle point integration.

In a similar way one can find the asymptotic form for $\langle|v_1-v_2|^\delta v_1^{2n}\rangle$ for large $n$.
Consider the equality,
\bea
\langle|v-v_1|^\delta v^{2n}\rangle=\int dvdv_1P(v)P(v_1)|v-v_1|^\delta v^{2n}.
\label{eq appdx 8}
\eea
As we are interested in the large $n$, the moments are dominated by values of $|v|$.
In this limit it is possible to approximate, $|v-v_1|\approx|v|$, and \eref{eq appdx 8} reduces to 
\bea
\langle|v-v_1|^\delta v^{2n}\rangle\sim\int dvP(v)|v|^{2n+\delta}
\label{eq appdx 9}.
\eea
Doing a transformation, $v=n^{1/\beta}t$ as before, we obtain
\bea
\langle|v-v_1|^\delta v^{2n}\rangle\sim n^{\frac{1+\delta}{\beta}}\int dte^{nf(t)+\Psi(n^{1/\beta}t)}n^{2n/\beta}
\label{eq appdx 10}
\eea
where,
\begin{align}
f(t)=-at^\beta+2\ln t
\label{eq appdx 11}
\end{align}
Doing a saddle point integration with respect to $t$, by maximizing $f(t)$, one obtains,
\bea
\langle|v-v_1|^\delta v^{2n}\rangle\approx n^{\frac{1+\delta}{\beta}}\sqrt{\frac{2\pi}{f''(t^*)n}}{t^*}^\delta\left(\frac{2n}{ae\beta}\right)^{\frac{2n}{\beta}}
e^{\Psi(n^{\frac{1}{\beta}}t^*)}~~
\label{eq appdx 12}
\eea
where we have substituted, the maximal value of $f(t)$,
\begin{align}
f(t^*)=\frac{2}{\beta}\ln\left(\frac{2}{ae\beta}\right),
\label{eq appdx 13}
\end{align}
and
\bea
t^*=\left(\frac{2}{a\beta}\right)^\frac{1}{\beta}.
\label{eq appdx 14}
\eea

The same calculation may be used to determine the asymptotic behavior of the noise distribution.
Considering the noise distribution with the form as in \eref{eq:noise-pdf},
\bea
\Phi(\eta)\sim |\eta|^{\tilde{\chi}}e^{-b|\eta|^\gamma} \text{for} |\eta|^2 \gg \langle \eta^2\rangle_\Phi, 
\label{eq appdx 15}
\eea
then
\be
N_{2n} = \langle \eta^{2n}\rangle \approx n^{\frac{1+\tilde{\chi}}{\gamma}}\sqrt{\frac{2\pi}{g''(y^*)n}}{y^*}^{\tilde{\chi}} \left[\frac{2 n}{b e \gamma}\right]^{\frac{2 n}{\gamma}}. 
\label{eq appdx 21}
\ee
where,
\begin{align}
g(y)=-by^\gamma+2\ln y
\label{eq appdx 22}
\end{align}
 and
\bea
y^*=\left(\frac{2}{b\gamma}\right)^\frac{1}{\gamma}.
\label{eq appdx 23}
\eea

\section{\label{appendix2} Taylor expansion of the  terms in Eq.~(\ref{eq:evolution}) arising from driving}

In this appendix, we analyse the terms [the last two terms] of Eq.~(\ref{eq:evolution}) arising from driving -- which we denote by $I$ -- for $r_w=1$:
\be
I=-\lambda_d P(v) +
\lambda_d \int \int d\eta  dv_1 \Phi(\eta) P(v_1) \delta\left[ - v_1+\eta -v \right],
\label{eq:a21}
\ee
by Taylor expanding the term in the integrand, and using the solution for the velocity distribution that we have obtained in the paper to identify the most dominant term in the expansion. Integrating over $v_1$, we obtain
\be
I=-\lambda_d P(v) +
\lambda_d \int d\eta  \Phi(\eta) P(v - \eta).
\label{eq:a22}
\ee
Taylor expanding $P(v-\eta)$ about $\eta=0$ and integrating over $\eta$, Eq.~(\ref{eq:a22}) reduces to
\be
I=\lambda_d \sum_{k=1}^\infty  \frac{\langle \eta^{2 k} \rangle}{(2 k)!} \frac{d^{2k}}{d v^{2k}} P(v)\equiv \sum_{k=1}^\infty t_k.
\label{eq:a23}
\ee
We now determine the dominant term in this expansion, given that $P(v)$ has the asymptotic behaviour as given in Eqs.~(\ref{velocity pdf with log correction 3})-(\ref{eq:last}).

From Eqs.~(\ref{velocity pdf with log correction 3})-(\ref{eq:last}), the velocity distribution has the asymptotic form 
\be 
\ln P(v) \sim -a |v|^\beta (\ln |v|)^\theta, \quad |v| \to \infty,
\label{eq:a24}
\ee
where $\theta>0$ for $\gamma >1$, $\delta>0$, and $\theta=0$ otherwise. Thus, 
\be
\frac{d^{2k}}{d v^{2k}} P(v) \approx \left[a \beta v^{\beta-1} (\ln |v|)^\theta\right]^{2 k} P(v), \quad |v| \to \infty.
\label{eq:a25}
\ee
The asymptotic form for the moments of $\eta$ is given in Eq.~(\ref{eq:noise-asym}). Using these asymptotic forms, we obtain the ratio of successive terms in Eq.~(\ref{eq:a23}) to be
\be
\frac{t_{k+1}}{t_k} \approx  \left( \frac{2 }{b \gamma}\right)^{\frac{2}{\gamma}} \frac{\left[a \beta v^{\beta-1} (\ln |v|)^\theta\right]^{2} }{4} k^{\frac{2(1-\gamma)}{\gamma}}.
\label{eq:a26}
\ee
We now analyse Eq.~(\ref{eq:a26}) for different cases.

Case I -- $\delta\leq 0$:  From Eq.~(\ref{velocity pdf with log correction 3}), we know that $\theta=0$ and $\beta =\min\left[\frac{2 + \delta}{2}, \gamma \right]$.
When $\delta<0$, then $\beta<1$. Thus, for large $v$, the ratio in Eq.~(\ref{eq:a26}) is much smaller than $1$ for small $k$. Also, for $\gamma >1$, the ratio decreases with $k$. Thus, the truncation at the first term is valid, and the expression for $\beta$ may be obtained from the continuum equations. However, when $\gamma <1$, it is possible that the expansion may break down as there could be contribution from large derivatives. Indeed, from comparing with our analytic solution, for $\gamma < 1+\delta/2$, the Taylor expansion about small $\eta$ breaks down. When $\delta=0$, then $\beta=1$. For $\gamma >1$, the most dominant term in the expansion, determined by $t_{k+1}/t_k = 1$, is a $k^*$ that is independent of $v$. Then, it is straightforward to show that $\beta=1$ is a consistent solution. Thus, while the dominant term in the expansion is not the first term, the result obtained from the continuum equations do not change.

Case II -- $\delta > 0$: Now, for $\gamma>1$, the solution of $t_{k^*+1}/t_{k^*} =1$ is given by $k^* \propto (\ln |v|)^{\frac{\theta \gamma}{\gamma-1}}$. Now, $k^*$ depends on $v$,  and it is clear that truncating the Taylor expansion at the first term will lead to wrong results. In fact, truncating at the first term gives $\beta = (2+\delta)/2$, while we have derived, without making any approximations, $\beta=1$.

\bibliographystyle{unsrt}

\end{document}